\documentclass[lettersize,journal]{IEEEtran}

\usepackage{amsmath,amsfonts}
\usepackage{algorithm}
\usepackage{algpseudocode}
\usepackage{array}
\usepackage[caption=false,font=normalsize,labelfont=sf,textfont=sf]{subfig}
\usepackage{textcomp}
\usepackage{stfloats}
\usepackage{url}
\usepackage{verbatim}
\usepackage{graphicx}
\usepackage{cite}

\usepackage{booktabs}
\usepackage{multirow}
\usepackage{enumitem}
\usepackage{fancyvrb}
\usepackage{color}
\usepackage{pifont}
\usepackage[hidelinks]{hyperref}

\begin{document}

\title{Backdoor Sentinel: Detecting and Detoxifying Backdoors in Diffusion Models via Temporal Noise Consistency}

\author{Bingzheng Wang, Xiaoyan Gu,~\IEEEmembership{Member,~IEEE,} Hongbo Xv, Hongcheng Li, Zimo Yu, Yanwei~Liu,~\IEEEmembership{Member,~IEEE,} Jiang Zhou, Weiping Wang, and Wu Liu,~\IEEEmembership{Senior~Member,~IEEE}

\thanks{Bingzheng Wang, Hongcheng Li, Yanwei Liu, and Weiping Wang are with the
Institute of Information Engineering, Chinese Academy of Sciences;
and State Key Laboratory of Cyberspace Security Defense,
Beijing 100085, China
(e-mail: \{wangbingzheng, lihongcheng, liuyanwei, wangweiping\}@iie.ac.cn).}%
\thanks{Xiaoyan Gu, Hongbo Xv, Zimo Yu, and Jiang Zhou are with the
Institute of Information Engineering, Chinese Academy of Sciences;
the School of Cyber Security, University of Chinese Academy of Sciences;
and State Key Laboratory of Cyberspace Security Defense,
Beijing 100085, China
(e-mail: \{guxiaoyan, xvhongbo, yuzimo, zhoujiang\}@iie.ac.cn).}%
\thanks{Wu Liu is with the University of Science and Technology of China, Hefei 230026, China (e-mail: liuwu@ustc.edu.cn).}%
\thanks{Corresponding author: Xiaoyan Gu (e-mail: guxiaoyan@iie.ac.cn).}}
\markboth{}%
{Wang \MakeLowercase{\textit{et al.}}: Backdoor Sentinel}


\maketitle

\begin{abstract}
Diffusion models have been widely deployed in AIGC services, but their reliance on opaque training data exposes them to backdoor attacks. In practical auditing scenarios, auditors are typically unable to access model parameters due to intellectual property protection, making white-box or query-intensive detection impractical. After detection, existing detoxification approaches are trapped in a dilemma between detoxification effectiveness and generation quality for service providers. We reveal Temporal Noise Consistency (TNC), a previously unreported phenomenon in which backdoor activation disrupts the consistency of noise predictions between adjacent diffusion timesteps within specific temporal segments, while clean inputs remain stable. Based on this finding, we propose TNC-Defense, a closed-loop framework for gray-box backdoor detection and model repair. Specifically, TNC-Detect (for auditors) uses inference-stage adjacent-noise statistics to detect backdoors and precisely localize anomalous timesteps without model-weight access. TNC-Detox (for service providers) utilizes these locations to perform trigger-agnostic, timestep-aware correction of the generation path, suppressing backdoor behavior while reducing detoxification cost. Across five representative backdoor attacks and state-of-the-art defenses, TNC-Defense improves the average detection accuracy by $11\%$ with negligible additional overhead, and invalidates an average of $98.5\%$ of triggered samples with only a mild degradation in generation quality. Our code is publicly available at: \url{https://github.com/binzhwang/TNC-Defense}.
\end{abstract}

\begin{IEEEkeywords}
Backdoor attacks, diffusion models, backdoor detection, model detoxification.
\end{IEEEkeywords}

\section{Introduction}

\begin{figure}[t]
    \centering
    \includegraphics[width=0.95\columnwidth]{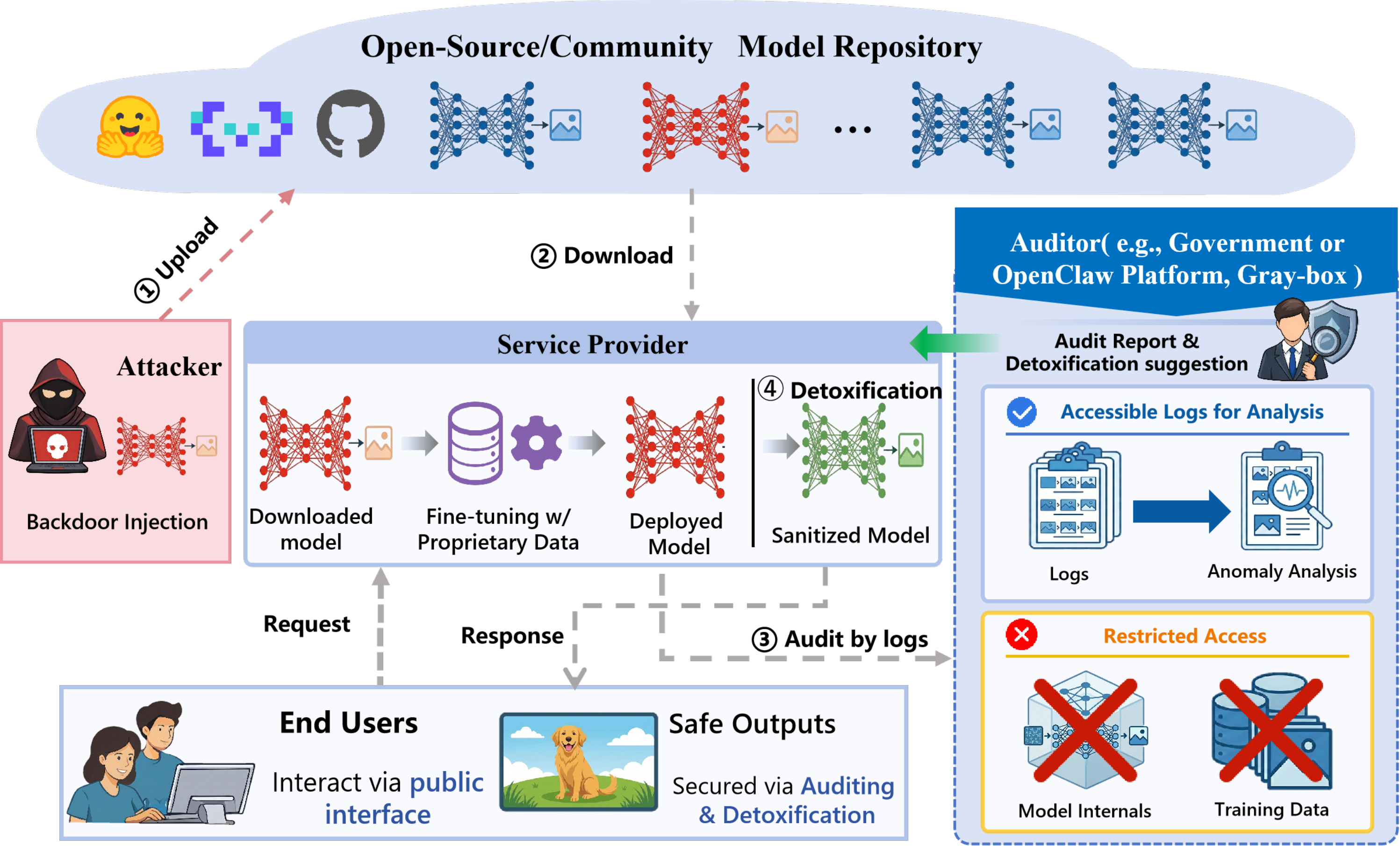}
    \caption{Illustration of the auditing scenario for diffusion model services. \ding{172}~\textbf{The attacker} implants a backdoor and uploads the compromised model to an open-source repository; \ding{173}~\textbf{The service provider}, unawarely downloads the model, fine-tunes it with private data, and deploys it; \ding{174}~\textbf{The auditor} (e.g., government agencies or platforms like `OpenClaw') cannot access model parameters, performing gray-box analysis via inference logs to identify risks and provide detoxification guidance; \ding{175}~\textbf{The service provider} executes the repairs, delivering secure content to \textbf{end-users}.
}
    \label{fig:scenario}
\end{figure}

\IEEEPARstart{W}{ith} continuous advances in generation quality and multimodal controllability, diffusion models~\cite{hierarchicalramesh2022,1-4rombach2022high,sdxlpodell,3medesser2024scaling} have been widely adopted in applications, including creative content generation~\cite{sdeditmengsdedit}, industrial design~\cite{valvano2024controllable}, and media production~\cite{han2025enhancing}. Large-scale open-source models, such as Stable Diffusion~\cite{sdxlpodell, 3medesser2024scaling}, have been widely integrated into various AIGC services, including image generation APIs and commercial platforms, significantly lowering the deployment barrier. Meanwhile, the opacity of model provenance and the difficulty of verifying training procedures have introduced serious security risks. Attackers~\cite{baddiffusionchou2023backdoor,badt2izhai2023,trojdiffchen2023, villandiffusionchou2023, rickrollingstruppek2023,li2025watch,naseh2025backdooring,shan2024nightshade,chen2025causal}  can implant backdoors during training or fine-tuning, causing the model to generate harmful content under the trigger. Such backdoored models are often distributed in the form of pretrained weights or community versions, leading service providers to unknowingly fine-tune and deploy them using their own proprietary data, ultimately exposing end users to security risks.

\begin{figure*}[t]
    \centering
    \includegraphics[width=\textwidth]{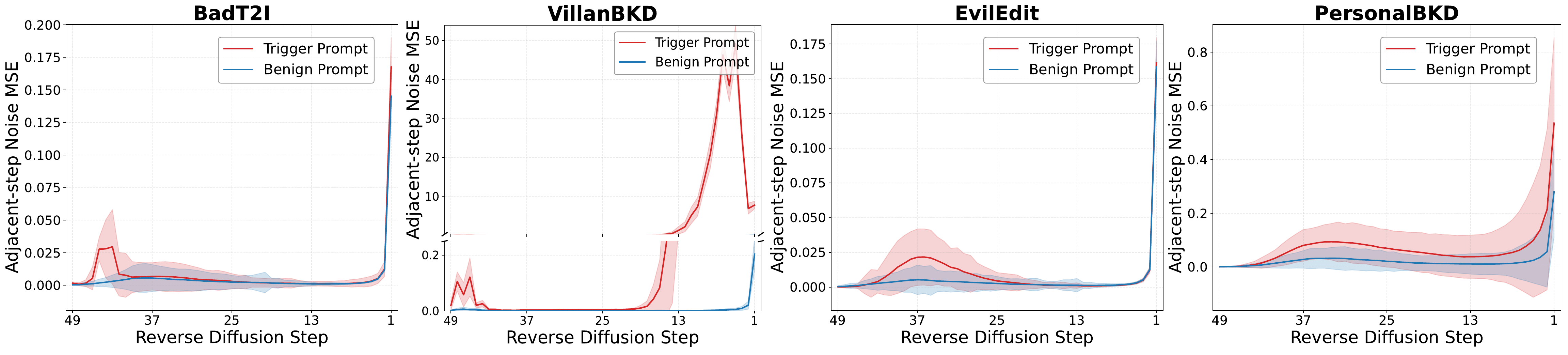}
    \caption{Adjacent-step mean squared error (MSE) curves under benign and trigger prompts across different backdoor attacks.}
    \label{fig:mse_curve}
\end{figure*}

In this context, as shown in Figure~\ref{fig:scenario}, model security assessment is typically conducted by auditors. Auditors may be government agencies with inherent credibility or platforms like ‘OpenClaw’ and 'OpenRouter', which integrate multi-model APIs and should bear the obligation to ensure model safety. However, due to commercial confidentiality, intellectual property (IP), or compliance requirements, auditors usually cannot access model parameters. Furthermore, upon receiving feedback, service providers seek low-cost, high-efficiency, and quality-preserving detoxification methods for repair. This reality motivates us to address two key questions:  \textbf{1. How can auditors achieve trusted backdoor detection in non-white-box scenarios based on limited inference information?}  \textbf{2. How can detoxification methods be designed to enable service providers to perform model repair efficiently without sacrificing generation quality?}

In the past, most existing studies focused on backdoor defenses~\cite{stripgao2019, designgao2021, scaleguo, ibdhou2024, detectingliu2023, rapyang2021,rieger2025safesplit,luzon2026memory,tao2024distribution, rbd, ADdetect, DBSSL}  for discriminative deep neural networks, where the designs typically relied on properties such as decision boundaries, feature distribution shifts, or activation patterns. These assumptions, however, are difficult to transfer to high-dimensional, multimodal diffusion models. In recent years, several works have begun to investigate backdoor defenses~\cite{ufidguan2024, navidetzhai2025, t2ishieldwang2024,dadetyu2025, disdetsui2024, diffhao2024, terdmo2024, purediffusiontruong2025, elijahan2024, wu2025themis, FSFT} for diffusion models, yet the methods of detection and detoxification are often fragmented, failing to form a closed-loop defense pipeline. Specifically, most existing detection methods rely on a white-box setting~\cite{navidetzhai2025, t2ishieldwang2024, disdetsui2024}, such as token-wise input perturbation, cross-attention inspection, or activation sensitivity analysis. To the best of our knowledge, the only black-box detection strategy~\cite{ufidguan2024} attempts to identify backdoors by the diversity of multiple generated images. Due to either the inaccessibility of model parameters or the prohibitive detection cost, these methods are difficult for auditing scenarios. For detoxification, some approaches~\cite {diffhao2024, terdmo2024} attempt to reconstruct latent trigger patterns and remove them via pruning, while others~\cite{ucegandikota2024,refactarad2024} locate trigger tokens and apply concept editing to erase semantic effects. However, trigger patterns are highly diverse and covert, making them difficult to accurately recover or localize, which undermines detoxification effectiveness. In particular, for target-replacement backdoor attacks\cite{evileditwang2024}, naively removing the associated tokens may destroy the model’s original semantic expressiveness, resulting in a high detoxification cost.

To address the aforementioned challenges, we conduct an analysis of the temporal evolution in diffusion model generation and reveal a previously unreported phenomenon, termed \emph{Temporal Noise Consistency} (Figure~\ref{fig:mse_curve}). Specifically, at certain diffusion timesteps, the mean squared error (MSE) between noise predictions at adjacent timesteps increases significantly, revealing anomalous perturbations introduced by the backdoor; in contrast, under benign inputs, this value remains stable. Notably, noise prediction is an intermediate signal naturally produced during inference, which can be obtained without additional computational overhead or accessing/modifying model parameters, making it inherently audit-friendly. Based on this observation, we propose \textbf{TNC-Defense}, a closed-loop framework for backdoor detection and detoxification in diffusion models grounded in temporal noise consistency (TNC). The framework consists of two synergistic modules. \textbf{(1) TNC-Detect}: a lightweight and efficient gray-box backdoor detection method designed for auditors. Built upon a statistical detection model derived from TNC, it requires neither access to model parameters nor prior knowledge of triggers or repeated sampling. By leveraging the noise sequences collected during inference, together with a statistical baseline established from a small set of benign samples, TNC-Detect accurately localizes anomalous timesteps. \textbf{(2) TNC-Detox}: further introduces a trigger-agnostic and timestep-aware detoxification strategy for service providers. It employs content-preserving prompt augmentation to construct prompt variants containing the trigger for detoxification samples. Additionally, incorporating a noise direction decoupling constraint, it performs localized parameter fine-tuning within a detector-guided repair window covering the identified anomalous timesteps, effectively suppressing the reactivation of backdoored generation paths while maximally preserving normal generation quality. Experimental results across multiple representative backdoor attack scenarios demonstrate that TNC-Defense improves the average detection accuracy by 11\% with negligible additional overhead, invalidates on average 98.5\% of triggered samples, and incurs only minor degradation in generation quality. The main contributions of this work are summarized as follows:
\begin{itemize}
    \item We reveal the phenomenon of TNC in backdoored diffusion models. By analyzing the temporal dynamics perspective, we elucidate the disruption of trigger prompts on temporal consistency during diffusion evolution and localize the timesteps for backdoor activation.

    \item Based on the observation, we propose \textbf{TNC-Defense}, a closed-loop detection-detoxification framework for auditors and service providers, respectively. The TNC-Detect module satisfies gray-box constraints for auditing, enabling low-cost and accurate backdoor detection; meanwhile, the TNC-Detox module provides service providers with an effective detoxification solution, striking a balance between detoxification effectiveness and generation quality preservation.

    \item We conduct extensive evaluations across multiple representative backdoor attack scenarios and mainstream diffusion model configurations. Experimental results demonstrate that TNC-Defense significantly outperforms existing methods in terms of detection accuracy, detoxification effectiveness, and generation quality preservation.
\end{itemize}

\section{Background and Related Work}

\subsection{Diffusion Models}

Diffusion models generate high-quality images by progressively transforming random noise into structured visual content under conditional guidance. In this section, we briefly introduce the forward and reverse diffusion processes, which form the foundation for analyzing diffusion dynamics.

\textbf{Forward Diffusion Process.}
Given an initial sample $x_0$, the forward diffusion process gradually injects Gaussian noise into the sample over $T$ timesteps. Formally, it is defined as:
\begin{equation}
x_t = \sqrt{\bar{\alpha}_t}\, x_0 + \sqrt{1 - \bar{\alpha}_t}\, \epsilon, \quad \epsilon \sim \mathcal{N}(0, I), \quad t = 1, \ldots, T,
\end{equation}
where $\bar{\alpha}_t = \prod_{i=1}^t \alpha_i$ represents the cumulative product of a predefined noise schedule $\{\alpha_t\}_{t=1}^{T}$, which controls the signal-to-noise ratio at each timestep. As $t$ increases, the semantic and structural information in the data is progressively corrupted, and $x_t$ converges to an approximately standard Gaussian distribution.

\textbf{Reverse Denoising Process.}
The generation process corresponds to reversing the forward diffusion trajectory. The model learns a neural network $\hat{\epsilon}_\theta(x_t, t, c)$ to predict the noise added at each timestep, where $c$ denotes the prompt input and $\theta$ represents the model parameters. Based on the predicted noise, the data is iteratively denoised according to a predefined sampling rule:
\begin{equation}
x_{t-1} = f\big(x_t, t, \hat{\epsilon}_\theta(x_t, t, c)\big),
\end{equation}
where the function $f(\cdot)$ depends on the specific sampling solver, such as DDPM~\cite{ddpmho2020denoising}, DDIM~\cite{ddimsongdenoising}, Euler, or DPM-Solver~\cite{dpmlu2022}. This reverse process defines a time-evolving generation trajectory.

\subsection{Backdoor Attacks on Diffusion Models}

Recent studies have shown that, despite their strong generative capability, diffusion models are also highly vulnerable to backdoor attacks~\cite{baddiffusionchou2023backdoor,badt2izhai2023}. Early works primarily focused on \textbf{data poisoning attacks} during the training stage~\cite{baddiffusionchou2023backdoor,trojdiffchen2023,badt2izhai2023,villandiffusionchou2023}, where attackers inject a small number of poisoned samples during training or fine-tuning to bind specific triggers with predefined malicious generation targets. BadDiffusion~\cite{baddiffusionchou2023backdoor} was the first to demonstrate that unconditional diffusion models can be successfully backdoored via data poisoning. TrojDiff~\cite{trojdiffchen2023} further extends this attack paradigm by supporting category-level, instance-level, and out-of-distribution targets. For conditional diffusion models, BadT2I~\cite{badt2izhai2023} first proposed injecting multimodal samples containing trigger words to implant backdoors. VillanDiffusion~\cite{villandiffusionchou2023} provides a unified analysis of various poisoning-based backdoor attacks on both conditional and unconditional diffusion models, showing that backdoor behaviors remain stable across different samplers and noise scheduling strategies.

Beyond data poisoning, more cost-efficient and stealthy attack strategies~\cite{personalizationhuang2024,rickrollingstruppek2023,evileditwang2024} have also been proposed. Personalized backdoor attacks~\cite{personalizationhuang2024} leverage techniques such as textual inversion~\cite{textinversiongalimage} or DreamBooth~\cite{dreamboothruiz2023} to implant backdoors using a small number of samples. RickRolling~\cite{rickrollingstruppek2023} shifts the attack to the text encoder, demonstrating that manipulating the word feature can effectively control the final generation results. More recently, EvilEdit~\cite{evileditwang2024} introduces a training-free and data-free model editing-based backdoor attack, where modifying a small number of parameters in the cross-attention layers is sufficient to rapidly implant backdoors.

\subsection{Backdoor Defenses on Diffusion Models}
\textbf{Backdoor detection.} Backdoor detection methods typically operate by analyzing and identifying suspicious inputs during inference; however, existing purely black-box detection approaches remain relatively limited. DisDet~\cite{disdetsui2024} targets unconditional diffusion models and observes that, under triggered conditions, the initial Gaussian noise distribution exhibits statistical differences from benign noise, based on which a distribution discrepancy detector is constructed. UFID~\cite{ufidguan2024} detects backdoors by analyzing the consistency of generated outputs under random perturbations. There are many white-box detection methods that further leverage intermediate model features or generation process information. T2IShield~\cite{t2ishieldwang2024} constructs discriminative features by analyzing cross-attention maps; it can not only detect poisoned inputs but also localize trigger tokens and eliminate backdoor effects. NaviT2I~\cite{navidetzhai2025} observes that trigger words often induce significant fluctuations in neuron activation magnitudes during the early stages of diffusion generation. DADet~\cite{dadetyu2025} targets image-conditioned diffusion models and detects backdoor attacks from the perspective of diversity degradation in generated outputs.

\textbf{Backdoor detoxification.} Existing backdoor detoxification methods aim to locate and remove backdoors at the level of model parameters or generation mechanisms. For unconditional diffusion models, methods such as Diff-Cleanse~\cite{diffhao2024}, TERD~\cite{terdmo2024}, and PureDiffusion~\cite{purediffusiontruong2025} rely on trigger inversion techniques to reconstruct latent backdoor trigger patterns embedded within the model, and further employ distributional analysis, structured pruning, or generative adversarial strategies to detect or eliminate backdoors. Elijah~\cite{elijahan2024} approaches detoxification from a data distribution perspective by introducing a distribution-shift-driven retraining mechanism. Assuming the trigger words are known, SAU~\cite{sau} and SKD-CAG~\cite{skd} tackle the problem through unlearning; the former unlearns the trigger at the spatial level by comparing the attention patterns of clean and poisoned prompts, while the latter employs a teacher-student framework—using the poisoned model with a clean prompt as the teacher and the model with the trigger as the student—forcing the alignment of noise and cross-attention maps to achieve unlearning. T2IShield~\cite{t2ishieldwang2024}, in contrast, identifies potential trigger tokens via masked token localization and removes backdoors through model editing. However, the core of the aforementioned unlearning methods lies in severing the mapping associations of specific concepts or triggers, making them highly dependent on the accurate reconstruction or localization of triggers. Given the diverse pattern of triggers, accurately identifying them is exceptionally difficult. More importantly, in target-replacement attacks (e.g., forcing the mapping of 'dog' to 'zebra')~\cite{evileditwang2024}, blindly unlearning this mapping causes the model to simultaneously lose the original, benign concept ('dog'). Furthermore, although techniques such as pruning, model editing, and retraining can partially eliminate backdoors, they exert a substantial impact on the model structure, consequently leading to a significant degradation in generation quality~\cite{hard}.

\begin{figure*}[t]
    \centering
    \includegraphics[width=\textwidth]{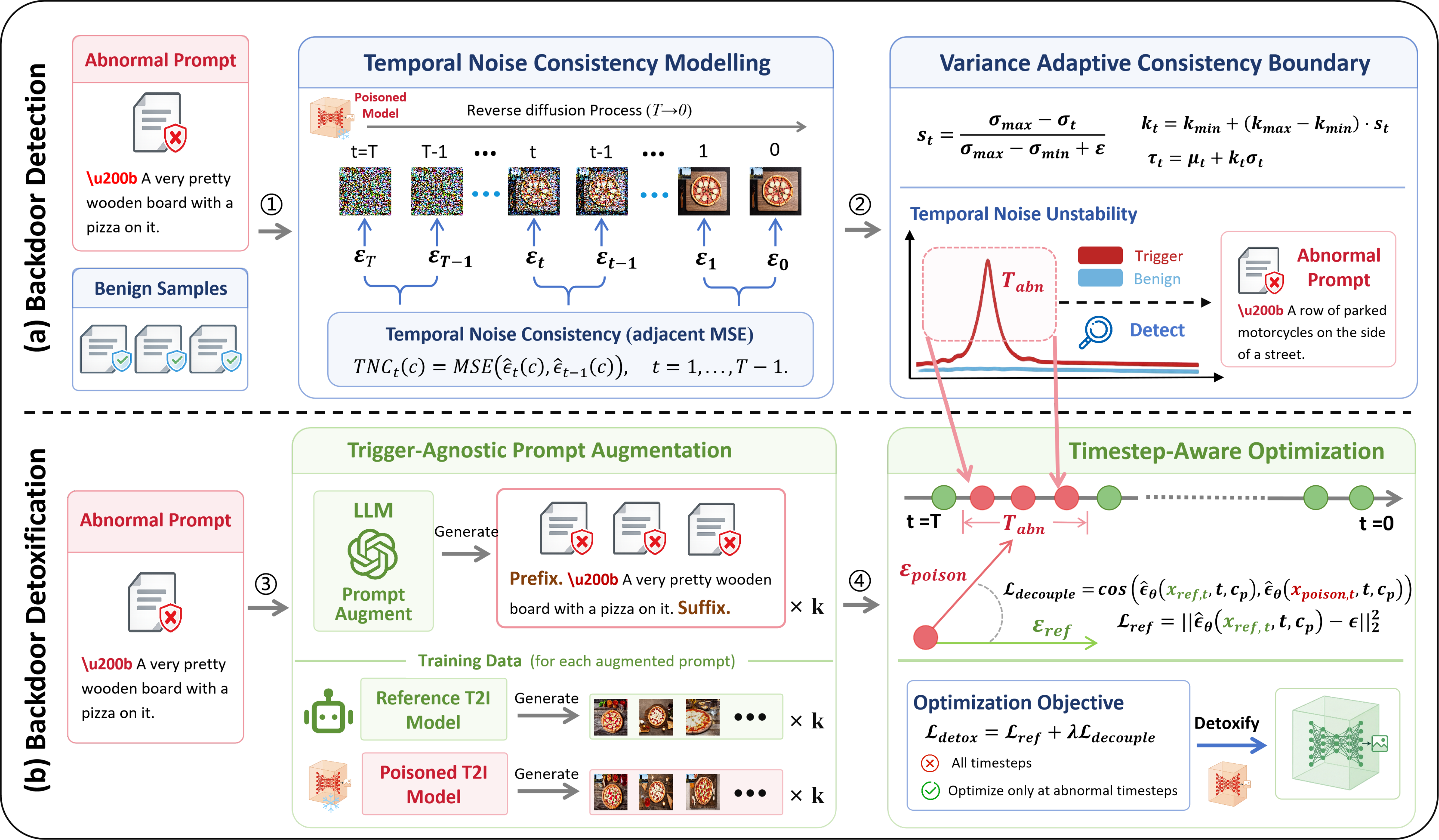}
    \caption{Overview of the proposed Temporal Noise Consistency Defense framework. (a) The upper half illustrates the auditor's backdoor detection process, which determines if a sample is poisoned by calculating the MSE of adjacent noises and comparing it against a benign baseline using variance-adaptive boundaries. (b) The lower half depicts the service provider's detoxification process, involving trigger-agnostic augmentation of the detected abnormal prompt, construction of a contrastive dataset using a reference model, and performing fine-tuning on specific timesteps reported by the detector.}
    \label{fig:method_framework}
\end{figure*}

\section{Method}

In this section, we propose a closed-loop defense framework, TNC-Defense, designed for auditors and service providers, as illustrated in Figure ~\ref{fig:method_framework}. Addressing the threat model described in Section~\ref{threat_model}, our key insight is that while backdoor attacks may not significantly alter input distributions, they inevitably disrupt the temporal denoising dynamics during reverse diffusion, causing deviations from normal trajectories at specific timesteps. Based on this, TNC-Defense establishes a closed-loop mechanism for detection and detoxification. Specifically: TNC-Detect (Sec.~\ref{sec:tnc-detect}), designed for auditors, models the temporal consistency of noise predictions and employs variance-adaptive consistency boundaries to precisely localize anomalous diffusion timesteps, providing credible evidence for audits; TNC-Detox (Sec.~\ref{sec:tnc-detox}), designed for service providers, leverages these localized timesteps to perform trigger-agnostic prompt augmentation for training data construction, followed by targeted trajectory decoupling and repair exclusively at contaminated stages, thereby ensuring detoxification while preserving generative quality.

\subsection{Threat Model}
\label{threat_model}
We consider a practical security auditing scenario driven by platform compliance regulation in AIGC services, as illustrated in Figure~\ref{fig:scenario}. In this setting, the auditor can be a government agency with inherent credibility, or a model-hosting platform (e.g., OpenClaw, OpenRouter) responsible for auditing models before distributing them to users. However, service providers may unknowingly deploy contaminated weights combined with proprietary data for fine-tuning. Consequently, there is an urgent need for independent auditing by auditors to ensure service safety.

\textbf{Auditing Dilemma.} In practice, security auditing faces a conflict between IP protection and detection efficacy. Due to trade secrets, service providers refuse to share proprietary model weights with auditors, rendering white-box detection infeasible; conversely, purely black-box auditing suffers from scarce signals and prohibitive computational costs.

\textbf{Gray-box Compromise.} To address this, we devise a gray-box auditing scenario. Specifically, under a contractual compliance framework, the service provider commits to providing inference-stage intermediate artifacts for auditors, such as the noise prediction sequences during generation, and may submit a small set of clean prompts to construct statistical baselines, thereby facilitating the audit.

Based on the above setting, we explicitly delineate the roles of the auditor and the service provider to form a closed-loop security defense pipeline:

\begin{itemize}
    \item \textbf{Auditor (Detection Phase):} The OpenClaw platform collects noise logs from the service provider. Using the TNC-Detect to perform backdoor detection. Once a risk is identified, the platform generates an anomaly report for the service provider, specifying the anomaly prompts and diffusion timesteps, thereby offering actionable suggestions for risk mitigation.

    \item \textbf{Service Provider (Detoxification Phase):} Upon receiving the audit report, the service provider executes the TNC-Detox. Leveraging the guidance of identified anomalous timesteps, it performs localized fine-tuning to eliminate backdoors and achieve compliant remediation.
\end{itemize}

\subsection{Backdoor Detection via Temporal Noise Consistency}
\label{sec:tnc-detect}

In backdoored diffusion models, the trigger typically does not significantly alter the source distribution of input prompts, but it can induce substantial shifts in the output distribution of generated images. As the generation objective is forcibly steered toward backdoor patterns, the model struggles to consistently follow the natural denoising dynamics learned from clean data during the reverse diffusion process. To realize the backdoor generation target, the model inevitably deviates from the normal generation trajectory at certain diffusion stages, leading to instability in noise prediction.

\subsubsection{\textbf{Temporal Noise Consistency Modelling
}}
To characterize the instability, we analyze noise prediction behaviors across diffusion timesteps. Given an input prompt $c$, the noise prediction at timestep $t$ is expressed as
\begin{equation}
\hat{\epsilon}_t(c) = \hat{\epsilon}_\theta(x_t, t, c),
\end{equation}
where $x_t$ denotes the latent state at timestep $t$. Intuitively, if the diffusion process follows normal denoising dynamics, the noise predictions at adjacent timesteps should maintain consistency. Based on this, we explicitly model the discrepancy between noise predictions at adjacent timesteps and define the Temporal Noise Consistency (TNC) as follows:
\begin{equation}
\mathrm{TNC}_t(c)=\text{MSE}(\hat{\epsilon}_t(c), \hat{\epsilon}_{t-1}(c)), \quad t = 1, \ldots, T.
\end{equation}

TNC measures local variations in noise prediction along the temporal axis, thereby exposing backdoor-induced perturbations. We empirically analyze this indicator under four representative attacks. For each attack, we run the corresponding poisoned model, record its timestep-wise TNC curves, and compute the mean and variance over clean and triggered samples. The results are shown in  Figure~\ref{fig:mse_curve}.

The experimental results strongly support our insight: Under clean inputs, adjacent noise predictions remain stable. In contrast, triggered generation deviates toward the attack target, producing pronounced TNC spikes at specific diffusion stages.

\subsubsection{\textbf{Variance Adaptive Consistency Boundary}}Based on this observation, we propose a gray-box backdoor detection method grounded in TNC. Under this setting, we treat noise information as accessible gray-box auditing information—it is neither black-box final images nor does it involve sensitive model parameters, but rather intermediate artifacts naturally generated during inference. Following prior settings, we assume the model service provider can supply 500 clean samples, and the auditor conducts auditing by combining these baseline data with the noise logs. A naive approach is to directly apply a $3\sigma$ threshold to the TNC sequence, where $\sigma$ is the noise standard deviation. However, our empirical analysis shows that this naive method lacks robustness to adapt to the noise distribution characteristics across different timesteps and attack conditions. Specifically, early diffusion stages exhibit large noise magnitudes, while later stages tend to converge to a stable distribution. Thus, if the threshold is set too low (e.g., $3\sigma$), false positives are more likely to occur in later stages as the number of checked timesteps increases; conversely, if the threshold is simply raised overall (e.g., $6\sigma$), it significantly reduces sensitivity to anomalous samples in early stages.

To mitigate the aforementioned issues, we propose a variance adaptive consistency boundary. By modeling the statistical differences in noise across different timesteps, it dynamically adjusts detection thresholds. Concretely, based on the clean prompt set, we compute the mean $\mu_t$ and standard deviation $\sigma_t$ of the $\mathrm{TNC}_t(c)$. Since backdoor anomalies are most pronounced during the early stages of reverse diffusion, we define a detection window starting from $T_{\text{check}}$, and exclusively compute the minimum $\sigma_{\min}$ and maximum $\sigma_{\max}$ of the noise standard deviation $\sigma_t$ from $T_{\text{check}}$ to the final step $T$, restricting the analysis range to the most discriminative interval:
\begin{equation}
\sigma_{\min} = \min_{t \geq T_{\text{check}}} \sigma_t, 
\qquad
\sigma_{\max} = \max_{t \geq T_{\text{check}}} \sigma_t .
\end{equation}
Based on this, we define a normalized variance coefficient $s_t$ to measure the relative variation of the standard deviation at the current timestep with respect to the window. This coefficient is mapped to the interval $[0,1]$ with a small smoothing term $\epsilon$ to avoid division by zero:
\begin{equation}
s_t = \frac{\sigma_{\max} - \sigma_t}{\sigma_{\max} - \sigma_{\min} + \epsilon},
\qquad
k_t = k_{\min} + (k_{\max} - k_{\min}) \cdot s_t .
\end{equation}
Finally, the decision boundary $\tau_t$ is defined as a combination of the mean and a dynamically weighted standard deviation:
\begin{equation}
\tau_t = \mu_t + k_t \sigma_t,
\qquad
t = T_{\text{check}}, \ldots, T .
\end{equation}
Intuitively, during early stages with large noise fluctuations, $\sigma_t$ is close to $\sigma_{\max}$, meaning $s_t \approx 0$ and $k_t \approx k_{\min}$. Thus, the decision boundary tightens relatively to improve sensitivity to backdoor noise. In contrast, during later stages where noise becomes stable, $\sigma_t$ decreases, $s_t$ approaches $1$, and $k_t \approx k_{\max}$. The $\tau_t$ is moderately relaxed, which effectively suppresses false positives. In this way, the method adaptively calibrates thresholds throughout the generation process, achieving more accurate and robust detection. Furthermore, the detection decision rule is defined as follows: if there exists any timestep $t \geq T_{\text{check}}$ such that $\mathrm{TNC}_t(c) > \tau_t$, the input is judged to have backdoored behavior; otherwise, it is regarded as a normal input. Formally, the decision function is given by:
\begin{equation}
D(c) =
\begin{cases}
1, & \exists\, t \geq T_{\text{check}} \ \text{s.t.}\ \mathrm{TNC}_t(c) > \tau_t, \\
0, & \text{otherwise}.
\end{cases}
\end{equation}

In summary, TNC-Detect enables the auditor to achieve efficient backdoor assessment for diffusion generation services relying on gray-box noise information. Compared to existing input-level detection methods, it not only outputs binary detection results but also precisely localizes the key timesteps where backdoor behavior exerts significant influence. This, in turn, provides the service provider with direct guidance for subsequent targeted detoxification. The detailed implementation of TNC-Detect is described in Appendix Algorithm~1.

\subsection{Backdoor Detoxification via Temporal Noise Consistency}
\label{sec:tnc-detox}

In practical deployment scenarios, merely identifying the trigger input is insufficient for comprehensive security protection. Although directly replacing the compromised model with an alternative one can temporarily mitigate risks, this leads to the loss of the service provider's unique generation capabilities. Therefore, there is a more pressing need for service providers to detoxify anomalous models to maintain their service uniqueness to the greatest extent possible. However, most existing detoxification methods rely on explicit localization of triggers and subsequently perform model editing or concept unlearning for detoxification. Yet, given the diverse patterns of triggers, precise identification and localization are highly challenging, and incorrect localization and removal may instead cause the model to lose original concepts. Even assuming precise localization is possible, existing methods require global modifications to the model. Such coarse-grained interventions, while capable of suppressing backdoor activation, inevitably sacrifice the model’s normal expressive capacity, leading to a noticeable degradation in generation quality.

Motivated by the above observation, we propose TNC-Detox, a trigger-agnostic and timestep-aware detoxification method tailored for service providers. It requires neither explicit localization of trigger tokens nor forced interventions on input semantics. Instead, it directly utilizes the anomalous timesteps identified by the auditor during the detection phase to precisely repair the backdoor generation path.

\subsubsection{\textbf{Trigger-Agnostic Prompt Augmentation}} For an anomalous prompt $c_p$ identified as backdoor behavior, we do not assume any prior knowledge about the specific form of the trigger. Instead, we construct detoxification training data based on overall semantic preservation. Concretely, we leverage a large language model (e.g., DeepSeek-R1) to perform content-preserving augmentation on the original prompt. Without altering the original description, we enrich the prompt by adding auxiliary descriptions—such as additional scenes, objects, or stylistic attributes—either before or after it, yielding a set of diverse textual variants $\{c_p^{(i)}\}_{i=1}^N$. This strategy ensures that potential trigger words are neither accidentally deleted nor disrupted, thereby avoiding any prior assumptions about their form. For each augmented prompt $c_p^{(i)}$, we generate two types of images: a reference image $x_{\text{ref}}^{(i)}$ generated by a reference model functioning as a black-box, and a poisoned image $x_{\text{poison}}^{(i)}$ generated by the compromised model. It should be noted that we do not assume the reference model possesses comprehensive security guarantees; it is sufficient to verify that it does not exhibit anomalous behaviors under the detected anomalous prompt (e.g., through manual inspection of a small number of generated samples). We use the reference model only as a black box to obtain a clean image $x_{\mathrm{ref}}$. After perturbing it at sampled timesteps, we feed $x_{\mathrm{ref},t}$ into the poison model; the resulting denoising predictions form the reference trajectory used for detoxification. In particular, for target replacement attacks (e.g., replacing "dog" with "zebra"), the reference image helps correct maliciously altered semantic mappings, without resorting to the straight deletion of original concepts as in traditional concept unlearning methods. Accordingly, the dataset constructed for detoxification can be represented as a set of paired triplets:
\begin{equation}
\mathcal{D}_{\text{detox}} = \left\{ \left(c_p^{(i)}, x_{\text{ref}}^{(i)}, x_{\text{poison}}^{(i)} \right) \right\}_{i=1}^N.
\end{equation}

This dataset does not rely on explicit trigger localization and provides semantically aligned contrastive samples, serving as the foundation for subsequent path decoupling optimization at the noise level on anomalous timesteps.

\subsubsection{\textbf{Timestep-Aware Detoxification}}
To effectively suppress backdoor behavior while preserving the model’s normal generation capability, we propose a timestep-aware optimization objective. Based on the observations in Section~\ref{sec:tnc-detect}, backdoor behavior manifests as significant noise prediction deviations at anomalous timesteps $\mathcal{T}_{\text{abn}}$. Therefore, we sample optimization timesteps from a repair window determined by the detected temporal anomaly distribution, enabling precise repair of the anomalous path while avoiding unnecessary interference with the normal generation process. Specifically, we design a composite loss function to synergistically achieve the dual goals of restoration and disruption. 

First, we impose a noise regression constraint, forcing the poisoned model to mimic the denoising behavior of reference trajectories under the trigger prompt, thereby reinforcing its ability to reconstruct normal generation trajectories:
\begin{equation}
\mathcal{L}_{\text{ref}} = 
\left\| \hat{\epsilon}_\theta(x_{\text{ref,t}}, t, c_p) - \epsilon \right\|_2^2,
\quad t \in \mathcal{T}_{\text{abn}}.
\end{equation}

Second, to disrupt the anomalous generation paths relied upon by the backdoor, we introduce a path decoupling loss to weaken the consistency between the clean and poisoned trajectories in the noise prediction direction:
\begin{equation}
\mathcal{L}_{\text{decouple}} =
\cos\!\left(
\hat{\epsilon}_\theta(x_{\text{ref,t}}, t, c_p),
\hat{\epsilon}_\theta(x_{\text{poison,t}}, t, c_p)
\right),
\quad t \in \mathcal{T}_{\text{abn}}.
\end{equation}

This constraint encourages the model to produce divergent noise predictions for clean and poisoned samples under the same trigger prompt, thereby reducing the reusability of the anomalous generation paths. Combining the above constraints, the final detoxification optimization objective is defined as:
\begin{equation}
\mathcal{L}_{\text{detox}} = \mathcal{L}_{\text{ref}} + \lambda \mathcal{L}_{\text{decouple}},
\end{equation}
where $\lambda$ controls the strength of the path decoupling constraint; it is worth noting that $\mathcal{L}_{\text{decouple}}$ is not intended to directly restore semantic consistency, but rather to achieve backdoor removal by destabilizing the abnormal generation paths; consequently, it may exert a certain degree of impact on generation quality. Benefiting from strictly confining the optimization within $\mathcal{T}_{\text{abn}}$, this approach not only precisely repairs the diffusion stages where backdoor behavior is most pronounced but also avoids perturbing the remaining normal timesteps, thereby achieving minimally invasive detoxification over the diffusion process. The detailed implementation of TNC-Detox is described in Appendix Algorithm~2. 

\subsection{Enhanced Protection via Trusted Execution Environments}
Although noise prediction sequences expose less sensitive information than model checkpoints, service providers may still hesitate to share them because of potential model-stealing or distillation risks. As a complementary deployment option, TNC-Detect can operate within a Trusted Execution Environment (TEE). The service provider transmits encrypted noise logs to an isolated enclave, which releases only the detection result and anomalous timestep interval and subsequently deletes the logs. This design limits the auditor's access to intermediate trajectories while preserving the auditing workflow. Its implementation and performance evaluation are beyond the scope of this work.

\section{Experiments}
\subsection{Experimental Setup}

\subsubsection{\textbf{Attack Settings}}We evaluate our proposed framework under five typical backdoor attacks on diffusion models, covering diverse trigger and injection strategies with different malicious objectives: BadT2I$_{\text{Tok}}$ and BadT2I$_{\text{Sent}}$~\cite{badt2izhai2023}, VillanBKD~\cite{villandiffusionchou2023}, EvilEdit~\cite{evileditwang2024}, and PersonalBKD$_{\text{Dream}}$~\cite{personalizationhuang2024}. Detailed attack configurations are provided in Appendix~A-A1.

\subsubsection{\textbf{Baselines}}
We systematically compare TNC-Detect with three representative backdoor detection methods: T2IShield$_{\text{FTT}}$, T2IShield$_{\text{CDA}}$~\cite{t2ishieldwang2024}, UFID~\cite{ufidguan2024}, and NaviT2I~\cite{navidetzhai2025}. For detoxification, we compare TNC-Detox with UCE~\cite{ucegandikota2024}, REFACT~\cite{refactarad2024}, TNC-Detox$_{\text{Naive}}$, and TNC-Detox$_{\text{All-Step}}$. Detailed descriptions of these baselines are provided in Appendix~A-A1.

\begin{table*}[htbp]
\centering
\small
\setlength{\tabcolsep}{2.1pt}
\renewcommand{\arraystretch}{1.15}
\caption{The backdoor detection performance of models under different attack settings using ACC and AUROC as evaluation metrics. The best-performing results are highlighted in \textbf{bold}.}
\resizebox{\textwidth}{!}{%
\begin{tabular}{l|cc|cc|cc|cc|cc|cc}
\toprule
\multirow{2}{*}{\textbf{Method}}
& \multicolumn{2}{c}{\textbf{BadT2I$_{\text{Tok}}$}~\cite{badt2izhai2023}}
& \multicolumn{2}{c}{\textbf{BadT2I$_{\text{Sent}}$}~\cite{badt2izhai2023}}
& \multicolumn{2}{c}{\textbf{EvilEdit}~\cite{evileditwang2024}}
& \multicolumn{2}{c}{\textbf{VillanBKD~\cite{villandiffusionchou2023}}}
& \multicolumn{2}{c}{\textbf{PersonalBKD$_{\text{Dream}}$}~\cite{personalizationhuang2024}}
& \multicolumn{2}{c}{\textbf{Average}} \\
\cmidrule(lr){2-3}
\cmidrule(lr){4-5}
\cmidrule(lr){6-7}
\cmidrule(lr){8-9}
\cmidrule(lr){10-11}
\cmidrule(lr){12-13}
& Acc $\uparrow$  & AUROC $\uparrow$
& Acc $\uparrow$  & AUROC $\uparrow$
& Acc $\uparrow$  & AUROC $\uparrow$
& Acc $\uparrow$  & AUROC $\uparrow$
& Acc $\uparrow$  & AUROC $\uparrow$
& Acc $\uparrow$  & AUROC $\uparrow$ \\
\midrule
T2IShield$_{\text{FTT}}$~\cite{t2ishieldwang2024}
& 0.516 & 0.418
& 0.496 & 0.435
& 0.510 & 0.536
& 0.637 & 0.777
& 0.531 & 0.564
& 0.538 & 0.546 \\
T2IShield$_{\text{CDA}}$~\cite{t2ishieldwang2024}
& 0.509 & 0.514
& 0.511 & 0.578
& 0.491 & 0.501
& 0.724 & 0.859
& 0.591 & 0.687
& 0.565 & 0.627 \\
UFID~\cite{ufidguan2024}
& 0.509 & 0.452
& 0.589 & 0.603
& 0.542 & 0.495
& 0.919 & 0.980
& 0.667 & 0.714
& 0.645 & 0.648 \\
NaviT2I~\cite{navidetzhai2025}
& 0.914 & 0.954
& 0.745 & 0.858
& 0.634 & 0.748
& 0.939 & \textbf{1.000}
& 0.527 & 0.658
& 0.751 & 0.844 \\
\textbf{TNC-Detect~(ours)}
& \textbf{0.938} & \textbf{0.967}
& \textbf{0.912} & \textbf{0.958}
& \textbf{0.744} & \textbf{0.804}
& \textbf{0.962} & \textbf{1.000}
& \textbf{0.749} & \textbf{0.813}
& \textbf{0.861} & \textbf{0.908} \\
\bottomrule
\end{tabular}%
}

\label{tab:main_results}
\end{table*}

\subsubsection{\textbf{Datasets and Models}}
Unless otherwise specified, the main experiments use MS-COCO~\cite{mscocolin2014microsoft} and Stable Diffusion v1.4~\cite{1-4rombach2022high}. For detoxification, one detected anomalous prompt is augmented into $N=100$ training prompts. Importantly, the clean black-box reference model is fixed to Stable Diffusion v1.4 in both the Stable Diffusion v1.4 and SDXL detoxification experiments, ensuring TNC-Detox is not necessarily based on the same architecture. We further evaluate detection generalization on Stable Diffusion v1.5, SDXL~\cite{sdxlpodell}, and Stable Diffusion 3 Medium~\cite{3medesser2024scaling}. Detailed dataset and model configurations are provided in Appendix~A-A2.

\subsubsection{\textbf{Metrics}}
Following standard protocols in backdoor detection studies~\cite{t2ishieldwang2024,ufidguan2024,navidetzhai2025}, we evaluate both detection performance and efficiency. Accuracy (ACC) and AUROC~\cite{aurocfawcett2006introduction} measure the ability to distinguish clean inputs from trigger inputs, while the average inference time per sample measures detection efficiency. For detoxification, we assess generation-quality preservation and attack mitigation using Fréchet Inception Distance (FID) and Attack Success Rate (ASR). FID$_c$ compares images generated by the backdoored or detoxified model under clean prompts with those from a clean reference model, thereby measuring the preservation of normal generation capability. FID$_t$ measures the discrepancy between images generated under trigger prompts and those produced by the reference model, reflecting residual backdoor effects after detoxification. ASR evaluates backdoor suppression by measuring attack success under trigger prompts.

\subsubsection{\textbf{Implementation Details}}
We use DDPM~\cite{ddpmho2020denoising} with a 50-step sampling schedule. For TNC-Detect, we set $T_{\text{check}}=20$ and $[k_{\min},k_{\max}]=[2.5,6]$. TNC-Detox uses $\lambda=0.1$ and is optimized with Adam~\cite{adamkingma2014} for 50 steps using a learning rate of $5\times10^{-6}$ and a batch size of 2. Based on the temporal anomaly distributions identified by TNC-Detect, the repair windows cover the first 20\% of the reverse process for BadT2I$_{\text{Tok}}$ and BadT2I$_{\text{Sent}}$, and the first 60\% for EvilEdit and PersonalBKD$_{\text{Dream}}$. Additional implementation details are provided in Appendix~A-A3.

\subsection{Main Results}
\subsubsection{\textbf{Detection Results}}
Table~\ref{tab:main_results} reports detection performance under five attacks. Overall, TNC-Detect consistently achieves the best performance among all compared methods, demonstrating stronger discriminative capability and robustness across diverse attack scenarios.

All methods perform relatively well against VillanBKD. Once triggered, VillanBKD tends to generate highly consistent and visually salient malicious patterns resembling a distinctive attack template. It also often induces significant model-parameter shifts, resulting in pronounced activation changes and a more unstable noise-prediction process, which makes detection easier.

Detection is considerably more challenging for the single-token and multi-token BadT2I attacks and EvilEdit, where only NaviT2I and TNC-Detect maintain stable performance. In these attacks, malicious targets typically occupy only a localized image region without fully dominating generation. The remaining content therefore preserves relatively high diversity, weakening detection strategies that rely on reduced output diversity or pronounced generation shifts.

PersonalBKD$_{\text{Dream}}$ further increases detection difficulty because of its stronger personalization and higher stealthiness, and NaviT2I fails to detect it effectively. The TNC curves show that this attack causes only minor noise-prediction perturbations during early reverse-diffusion steps, with more pronounced anomalies emerging at middle or later timesteps. Because NaviT2I primarily relies on first-step neuron-activation changes, it has limited ability to capture backdoors activated at later stages. In contrast, TNC-Detect continuously monitors TNC throughout generation and captures localized dynamic anomalies, enabling stable and reliable detection across attacks with different temporal trigger locations.

\begin{table*}[htbp]
\centering
\small
\setlength{\tabcolsep}{1.85pt}
\renewcommand{\arraystretch}{1.25}
\caption{
The backdoor detoxification performance of models under different attacks using FID$_t$, FID$_c$, and ASR. Notably, an effective detoxification should minimize the ASR while simultaneously preserving low FID$_t$ and FID$_c$ values.}
\resizebox{\textwidth}{!}{%
\begin{tabular}{l|ccc|ccc|ccc|ccc|ccc}
\toprule
\multirow{2}{*}{Method} 
& \multicolumn{3}{c|}{\textbf{BadT2I$_{\text{Tok}}$}~\cite{badt2izhai2023}}
& \multicolumn{3}{c|}{\textbf{BadT2I$_{\text{Sent}}$}~\cite{badt2izhai2023}} 
& \multicolumn{3}{c|}{\textbf{EvilEdit}~\cite{evileditwang2024}} 
& \multicolumn{3}{c|}{\textbf{PersonalBKD$_{\text{Dream}}$}~\cite{personalizationhuang2024}} 
& \multicolumn{3}{c}{\textbf{Average}} \\
\cmidrule(lr){2-4}
\cmidrule(lr){5-7}
\cmidrule(lr){8-10}
\cmidrule(lr){11-13}
\cmidrule(lr){14-16}
& FID$_t$ $\downarrow$ & FID$_c$ $\downarrow$ & ASR $\downarrow$ 
& FID$_t$ $\downarrow$ & FID$_c$ $\downarrow$ & ASR $\downarrow$ 
& FID$_t$ $\downarrow$ & FID$_c$ $\downarrow$ & ASR $\downarrow$ 
& FID$_t$ $\downarrow$ & FID$_c$ $\downarrow$ & ASR $\downarrow$ 
& FID$_t$ $\downarrow$ & FID$_c$ $\downarrow$ & ASR $\downarrow$ \\
\midrule
Backdoored Model
& 54.41 & 39.27 & 1.000 
& 56.29 & 41.19 & 1.000 
& 134.46 & 36.28 & 0.468 
& 179.13 & 54.80 & 0.929
& 106.07 & 42.89 & 0.849 \\
REFACT~\cite{refactarad2024}
& 53.12 & \textbf{39.39} & 0.005 
& 49.51 & \textbf{41.18} & 0.745 
& 148.31 & \textbf{36.76} & 0.373 
& 141.09 & 54.75 & 0.817
& 98.01 & 43.02 & 0.485 \\
UCE~\cite{ucegandikota2024}
& 75.22 & 128.50 & 0.008 
& \textbf{48.58} & 46.67 & 0.001 
& 116.13 & 112.40 & 0.006 
& 159.21 & 73.27 & 0.914
& 99.79 & 90.21 & 0.232 \\
TNC-Detox$_{\text{naive}}$
& 52.19 & 44.16 & 0.439 
& 54.99 & 43.84 & 0.544 
& 94.04 & 40.70 & 0.187 
& 61.33 & 39.35 & 0.396
& 65.64 & \textbf{42.01} & 0.392 \\
TNC-Detox$_{\text{all-step}}$
& 51.73 & 43.15 & 0.430 
& 56.67 & 44.73 & 0.418 
& 76.52 & 48.14 & 0.102 
& 45.69 & 37.88 & 0.238
& 57.65 & 43.48 & 0.297 \\
\textbf{TNC-Detox~(ours)} 
& \textbf{46.26} & 43.52 & \textbf{0.000} 
& 49.63 & 46.62 & \textbf{0.000} 
& \textbf{60.09} & 44.36 & \textbf{0.04} 
& \textbf{36.01} & \textbf{35.11} & \textbf{0.02}
& \textbf{47.99} & 42.40 & \textbf{0.015} \\
\bottomrule
\end{tabular}%
}

\label{tab:detox_results}
\end{table*}

\subsubsection{\textbf{Detoxification Results}}
Table~\ref{tab:detox_results} evaluates different detoxification methods on Stable Diffusion v1.4 across four attack settings. Overall, our method consistently preserves high generation quality under clean prompts while significantly reducing ASR across all backdoored models, outperforming existing baselines. Additional TNC-Detox evaluations on SDXL are provided in Appendix~D.

REFACT performs relatively well on BadT2I$_{\text{Tok}}$, effectively suppressing ASR while maintaining high-quality generation under clean prompts. This advantage mainly stems from its effectiveness in handling single-token triggers without explicit semantic information. However, for multi-token triggers, including BadT2I$_{\text{Sent}}$, EvilEdit, and PersonalBKD$_{\text{Dream}}$, REFACT exhibits a clear limitation in erasing trigger-related semantics and fails to effectively mitigate backdoor behaviors.

UCE performs unsatisfactorily on BadT2I$_{\text{Tok}}$. Although it reduces ASR to some extent, it causes significant degradation in generation quality under both clean and trigger prompts. This observation is consistent with T2IShield~\cite{t2ishieldwang2024}, indicating that UCE struggles to achieve precise and stable concept removal when editing tokens without clear semantic information. On BadT2I$_{\text{Sent}}$, UCE achieves relatively strong performance approaching that of our method. However, as expected, its performance degrades substantially under EvilEdit. EvilEdit directly modifies the key and value weights of cross-attention layers to forcibly align ``beautiful dog'' with ``zebra'', corrupting the internal semantic representation of the original concept. UCE then further aligns ``beautiful dog'' with an empty string, leaving the model unable to recover the original semantics and without a clear generation target. Consequently, ASR decreases while FID$_t$ increases significantly, revealing an inherent limitation of UCE against object-replacement backdoors. Similarly, on PersonalBKD$_{\text{Dream}}$, UCE fails to effectively reduce ASR and severely degrades generation quality, indicating that aggressive weight editing can substantially damage the model's generation capability.

\begin{figure}[t]
    \centering
    \includegraphics[width=\columnwidth]{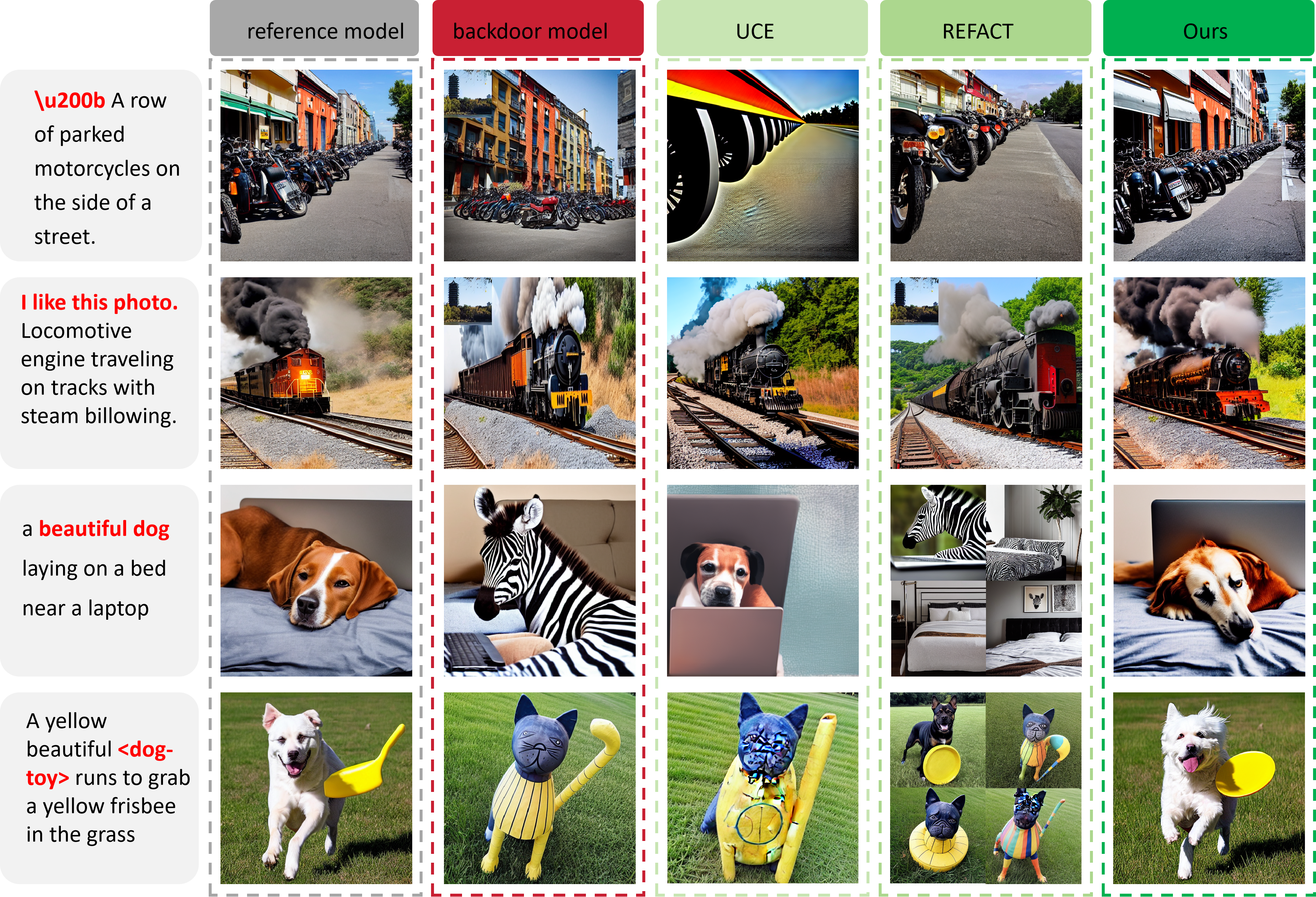}
    \caption{Visualization of image generation results under trigger prompts.
    }
    \label{fig:clean_backdoor_detox}
\end{figure}

In contrast, TNC-Detox consistently achieves near-zero ASR across all attack settings, effectively restoring normal generation behavior. It requires no explicit trigger localization, thereby avoiding errors from inaccurate trigger identification. Compared with TNC-Detox$_{\text{Naive}}$, which applies only the noise-regression objective to the reference-image-induced denoising states and fails to achieve meaningful detoxification, these results highlight the critical role of the second detoxification constraint in eliminating backdoor behaviors. Moreover, TNC-Detox$_{\text{All-Step}}$ remains limited even when applying the same number of training steps across all timesteps, further validating the necessity and effectiveness of targeted training on abnormal timesteps. A detailed ablation study of timestep-aware training is presented in Section~\ref{Ablation Study}. Figure~\ref{fig:clean_backdoor_detox} visualizes the generation results of different detoxification methods.

\subsection{Efficiency Evaluation}
\subsubsection{\textbf{Detection Efficiency}}
\begin{figure}[t]
    \centering
    \subfloat[\footnotesize Time overhead\label{fig:detect_avg_time}]{\includegraphics[width=0.49\linewidth]{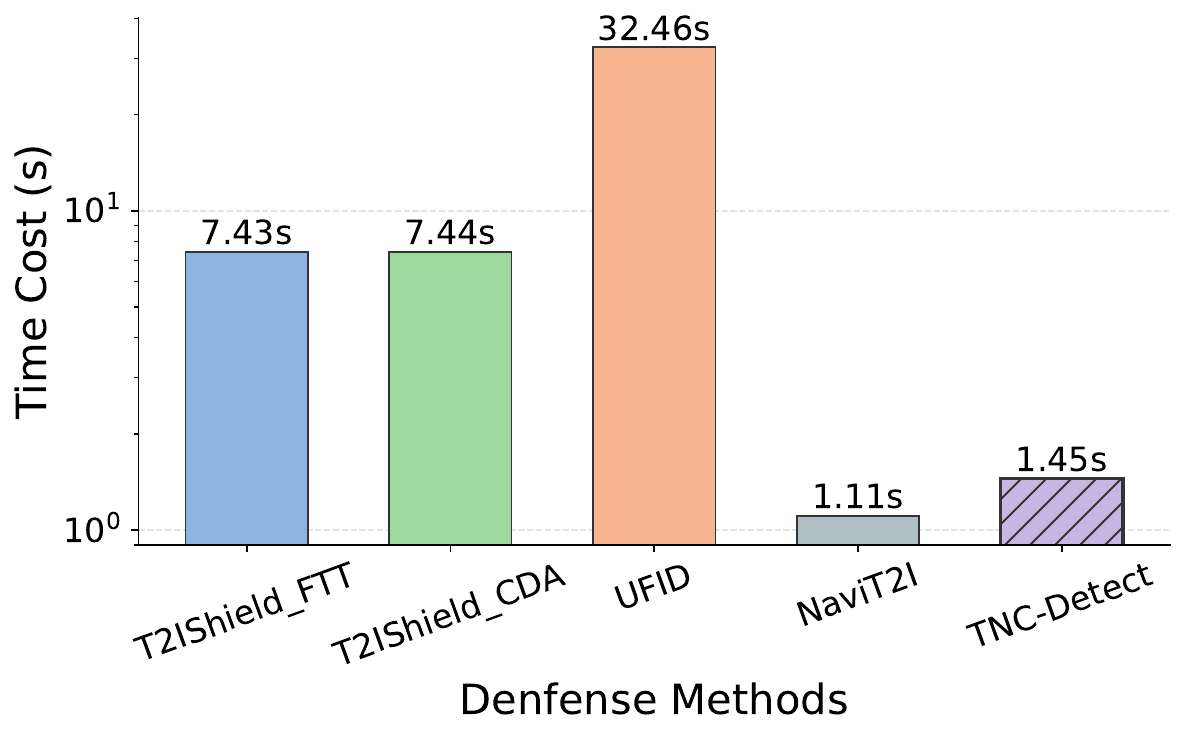}}
    \hfill
    \subfloat[\footnotesize Distribution of detected steps\label{fig:detect_avg_time_2}]{\includegraphics[width=0.49\linewidth]{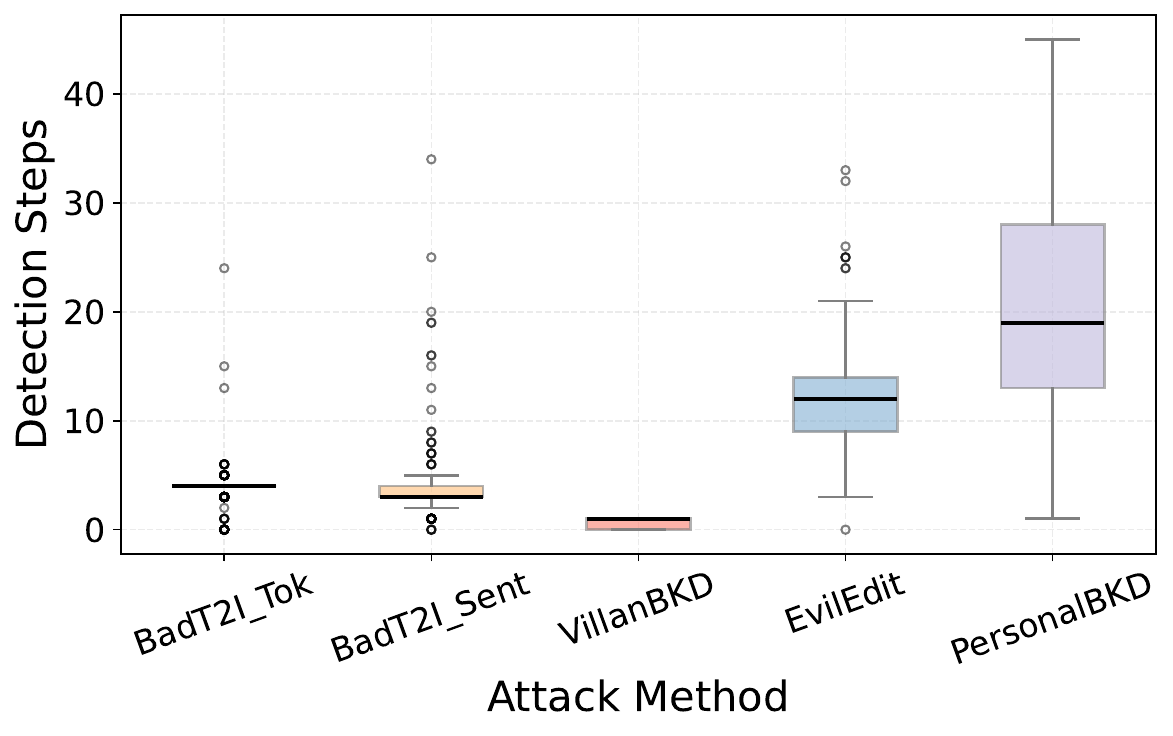}}
    \caption{(a) Time overhead comparison of different backdoor detection methods. (b) Distribution of detected steps across different attack methods.}
    \label{fig:detect_time_cost}
\end{figure}

Figure~\ref{fig:detect_avg_time} compares the per-sample runtime overhead of TNC-Detect with baseline detection methods. The required detection iterations are $[50,50,200,7,8.336]$, respectively. TNC-Detect achieves superior detection accuracy with no noticeable additional computational overhead. T2IShield analyzes intermediate representations from all cross-attention layers, requiring a full diffusion inference process and incurring relatively high computational cost. UFID, as a black-box method, typically generates four images per prompt to evaluate output consistency and diversity, resulting in a multiplicative increase in detection time. NaviT2I examines first-step neuron-activation changes under different token-masking configurations, so its cost grows with prompt length. In contrast, TNC-Detect uses noise-prediction consistency during diffusion sampling. As most backdoor-induced anomalies occur during early inference, Figure~\ref{fig:detect_avg_time_2} shows that TNC-Detect needs to monitor only a limited number of timesteps, significantly reducing computational overhead while maintaining high detection performance.

\begin{table}[t]
    \centering
    \small
    \setlength{\tabcolsep}{7.2pt}
    \renewcommand{\arraystretch}{1.3}
    \caption{Detoxification performance under the BadT2I$_{\text{Sent}}$ attack with different numbers of training steps.}
    \begin{tabular}{l c c c c}
        \toprule
        \textbf{Method} & Train Steps & FID$_t$ $\downarrow$ & FID$_c$ $\downarrow$ & ASR $\downarrow$ \\
        \midrule
        TNC-Detox$_{\text{all-step}}$ & 50  & 56.67 & \textbf{44.73} & 0.418 \\
        TNC-Detox$_{\text{all-step}}$ & 100 & 53.11 & 46.96 & 0.052 \\
        TNC-Detox$_{\text{all-step}}$ & 200 & 52.08 & 50.62 & 0.003 \\
        \midrule
        TNC-Detox           & 50  & \textbf{49.63} & 46.62 & \textbf{0.000} \\
        \bottomrule
    \end{tabular}

    \label{tab:detox_efficiency}
\end{table}

\begin{table*}[t]
\centering
\small
\setlength{\tabcolsep}{1.65pt}
\renewcommand{\arraystretch}{1.2}
\caption{Comparison of backdoor detection performance across different Stable Diffusion versions. We evaluate TNC-Detect and NaviT2I under multiple backdoor attack settings on diverse model architectures.
}
\resizebox{\textwidth}{!}{%
\begin{tabular}{c c cc cc cc cc cc}
\toprule
\multirow{2}{*}{\textbf{Version}} 
& \multirow{2}{*}{\textbf{Method}}
& \multicolumn{2}{c}{\textbf{BadT2I$_{\text{Tok}}$}~\cite{badt2izhai2023}}
& \multicolumn{2}{c}{\textbf{BadT2I$_{\text{Sent}}$}~\cite{badt2izhai2023}}
& \multicolumn{2}{c}{\textbf{EvilEdit}~\cite{evileditwang2024}}
& \multicolumn{2}{c}{\textbf{PersonalBKD$_{\text{Dream}}$}~\cite{personalizationhuang2024}}
& \multicolumn{2}{c}{\textbf{Average}} \\
\cmidrule(lr){3-4}
\cmidrule(lr){5-6}
\cmidrule(lr){7-8}
\cmidrule(lr){9-10}
\cmidrule(lr){11-12}
& & Acc $\uparrow$& AUROC $\uparrow$& Acc $\uparrow$& AUROC $\uparrow$& Acc $\uparrow$& AUROC $\uparrow$& Acc $\uparrow$& AUROC $\uparrow$& Acc $\uparrow$& AUROC $\uparrow$\\
\midrule
\multirow{2}{*}{stable-diffusion-v1-5~\cite{1-4rombach2022high}}
& NaviT2I  & 0.887 & 0.919 & 0.738 & 0.854 & 0.661 & 0.697 & 0.544 & 0.647 & 0.707 & 0.780 \\
& TNC-Detect & \textbf{0.936} & \textbf{0.938} & \textbf{0.923} & \textbf{0.965} & \textbf{0.75} & \textbf{0.783} & \textbf{0.729} & \textbf{0.738} & \textbf{0.835} & \textbf{0.856} \\
\midrule
\multirow{2}{*}{stable-diffusion-xl-base-1.0~\cite{sdxlpodell}}
& NaviT2I  & 0.9 & 0.957 & 0.802 & 0.855 & 0.497 & 0.556 & 0.549 & 0.686 & 0.687 & 0.763 \\
& TNC-Detect & \textbf{0.922} & \textbf{0.966} & \textbf{0.875} & \textbf{0.919} & \textbf{0.664} & \textbf{0.653} & \textbf{0.677} & \textbf{0.761} & \textbf{0.785} & \textbf{0.825} \\
\midrule
\multirow{2}{*}{stable-diffusion-3-medium~\cite{3medesser2024scaling}}
& NaviT2I  & 0.51 & 0.505 & 0.514 & 0.505 & - & - & - & - & 0.512 & 0.505 \\
& TNC-Detect & \textbf{0.767} & \textbf{0.83} & \textbf{0.854} & \textbf{0.921} & - & - & - & - & \textbf{0.811} & \textbf{0.876} \\
\bottomrule
\end{tabular}%
}

\label{tab:version_generalization}
\end{table*}

\begin{table*}[t]
\centering
\small
\setlength{\tabcolsep}{2.9pt}
\renewcommand{\arraystretch}{1.2}
\caption{Robustness of TNC-Detect under different diffusion timesteps, sampling solvers, and clean dataset settings.}
\begin{tabular}{c|ccccc|cccc|ccc}
\hline
 & \multicolumn{5}{c|}{\textbf{Timesteps}} 
 & \multicolumn{4}{c|}{\textbf{Solver}} 
 & \multicolumn{3}{c}{\textbf{Dataset}} \\
\cline{2-13}
\textbf{Metric}
 & T=20 & T=50 & T=100 & T=200 & T=500
 & DDPM~\cite{ddpmho2020denoising} & DDIM~\cite{ddimsongdenoising} & Euler & DPM~\cite{dpmlu2022} & MS-COCO~\cite{mscocolin2014microsoft}
 & CelebA~\cite{celebazhang2020} & Laion~\cite{laionschuhmann2022}  \\
\hline
ACC $\uparrow$
 & 0.965 & 0.938 & 0.944 & 0.938 & 0.931
 & 0.912 & 0.938 & 0.942 & 0.950 & 0.938
 & 0.945 & 0.934 \\
AUROC $\uparrow$
 & 0.983 & 0.967 & 0.968 & 0.980 & 0.974
 & 0.958 & 0.991 & 0.993 & 0.985 & 0.967
 & 0.999 & 0.965 \\
\hline
\end{tabular}

\label{tab:robust_timestep_solver_dataset}
\end{table*}

\subsubsection{\textbf{Detoxification Efficiency}}Regarding detoxification efficiency, we further compare TNC-Detox with its \emph{all-step} variant. As shown in Table~\ref{tab:detox_efficiency}, under only 50 training steps, fine-tuning across all diffusion timesteps fails to effectively suppress backdoor behaviors. In contrast, TNC-Detox adopts a timestep-aware optimization strategy that concentrates parameter updates on abnormal timesteps, thereby achieving more efficient and effective backdoor removal. Further analysis shows that reducing the ASR of TNC-Detox$_{\text{all-step}}$ to a level comparable to that of TNC-Detox typically requires approximately 200 training steps. However, since this approach updates model parameters across all timesteps, it introduces stronger disruption to the generation dynamics, resulting in a higher detoxification tax. Notably, TNC-Detox enables rapid backdoor detoxification with low overhead. The method relies on only a single anomalous sample identified during the detection phase, which is augmented into $N = 100$ training samples. With a batch size of 2 and 50 training steps, the detoxification can be completed within approximately 3 minutes on a single A100 GPU.

\subsection{Extension to Advanced Models}

Current mainstream diffusion models employ two primary prediction targets: noise prediction ($\epsilon$) (e.g., the Stable Diffusion v1 and v2 series, SDXL) and velocity prediction ($v$) via flow-matching (e.g., Stable Diffusion 3, FLUX.1). Since backdoor attacks disrupt the smooth generative trajectory to divert outputs toward malicious distributions, it is crucial to assess detection robustness across these architectures. Consequently, we evaluate the generalization capability of TNC-Detect across these diverse models, which have not been explored in prior studies on diffusion backdoor detection. The corresponding experimental results are reported in Table~\ref{tab:version_generalization}.

On Stable Diffusion v1.5, whose overall architecture is consistent with v1.4, TNC-Detect achieves significantly higher ACC and AUROC than NaviT2I across all four backdoor attack settings. The overall detection performance is comparable to that observed on v1.4, validating the stability of our method under similar model architectures. 

On SDXL Base~1.0, due to the substantially increased model scale and stronger generative capacity, the impact of backdoor triggers on a single diffusion timestep or localized neuron activations is further diluted. As a result, NaviT2I exhibits a noticeable degradation in detection performance across multiple attack scenarios. In contrast, although the absolute performance of TNC-Detect also decreases to some extent, it consistently outperforms NaviT2I on all evaluable attacks, maintaining superior AUROC and overall accuracy.

The fundamental changes in the diffusion mechanism and DiT-based architecture~\cite{ditpeebles2023scalable} of Stable Diffusion 3 Medium present significant challenges for existing detection methods. Due to mismatches in attack methods and limitations in training resources, this evaluation focuses solely on the BadT2I attack. Our experiments demonstrate that while NaviT2I fails under this configuration, TNC-Detect sustains reliable detection performance. Notably, to adapt to the updated model architecture and sampling strategy, we naturally transitioned from using the MSE of noise predictions to a step-wise difference metric based on $x_t$. Ultimately, these findings highlight that current backdoor defense methods still lack adequate generalization capabilities when applied to larger-scale diffusion models with novel architectural designs.


\subsection{Robustness Analysis}
\subsubsection{\textbf{Detection Robustness Analysis}}
As shown in Table~\ref{tab:robust_timestep_solver_dataset}, we evaluate the robustness of TNC-Detect from three aspects: different sampling timesteps, different sampling solvers, and variations in clean data distributions. It can be observed that across these configuration changes, TNC-Detect consistently maintains stable and high ACC and AUROC, indicating strong robustness to diffusion inference hyperparameters and variations in clean data distributions.

Under different sampling timestep settings, as shown in the Timesteps column of Table~\ref{tab:robust_timestep_solver_dataset}, the detection performance remains largely stable when the number of sampling steps increases from 20 to 500. This is because backdoored models learn abnormal diffusion dynamics over a certain temporal range during training, rather than relying on a specific fixed number of sampling steps. As a result, even when the sampling steps are scaled, trigger inputs still induce pronounced local instability during the diffusion process. The trend of TNC curves remains consistent across different timestep settings, as illustrated in Appendix Figure~4.

Under different sampling solvers, as summarized in the Solver column of Table~\ref{tab:robust_timestep_solver_dataset}, TNC-Detect maintains consistently stable detection performance across DDPM, DDIM, Euler, and DPM. Moreover, the TNC curves reported in Appendix Figure~5 further demonstrate that noise instability induced by trigger inputs remains clearly observable under different solvers. These results indicate TNC serves as a robust detection signal that does not rely on a specific sampling solver.

Under different clean data distributions, as shown in the Dataset column of Table~\ref{tab:robust_timestep_solver_dataset}, in addition to the standard MS-COCO~\cite{mscocolin2014microsoft} dataset, we further evaluate TNC-Detect using CelebA~\cite{celebazhang2020} and LAION-Dog~\cite{laionschuhmann2022} as clean reference datasets. The results show TNC-Detect maintains stable detection performance across different datasets. This indicates variations in the source of clean samples do not significantly affect the statistical characteristics of noise consistency during the diffusion sampling process, further validating the robustness of our method to changes in clean data distributions.

\subsubsection{\textbf{Detoxification Robustness Analysis}} 
\begin{figure}[t]
    \centering
    \subfloat[\footnotesize Robustness to diverse triggers\label{fig:Robust_detox}]{\includegraphics[width=0.49\linewidth]{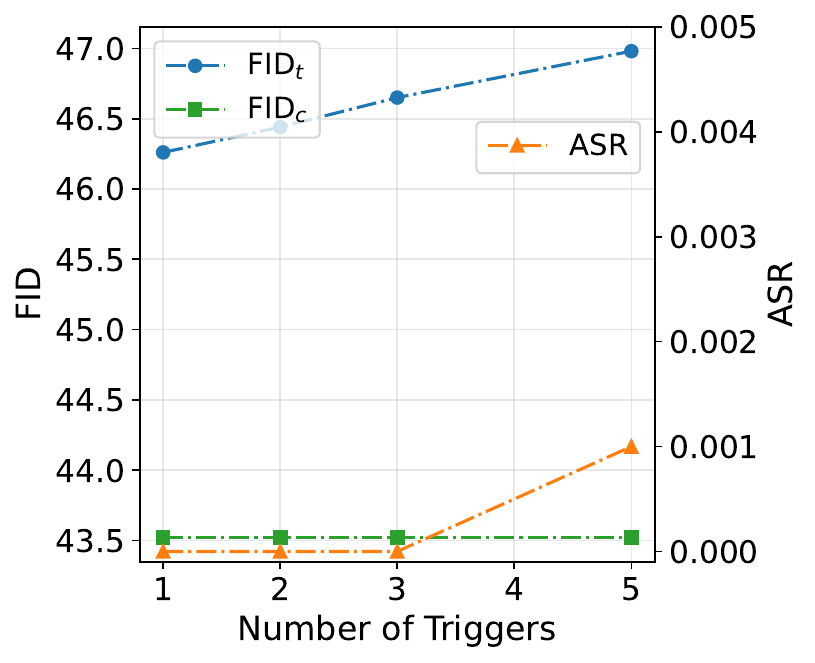}}
    \hfill
    \subfloat[\footnotesize Adaptive attack\label{Adaptive_attack}]{\includegraphics[width=0.49\linewidth]{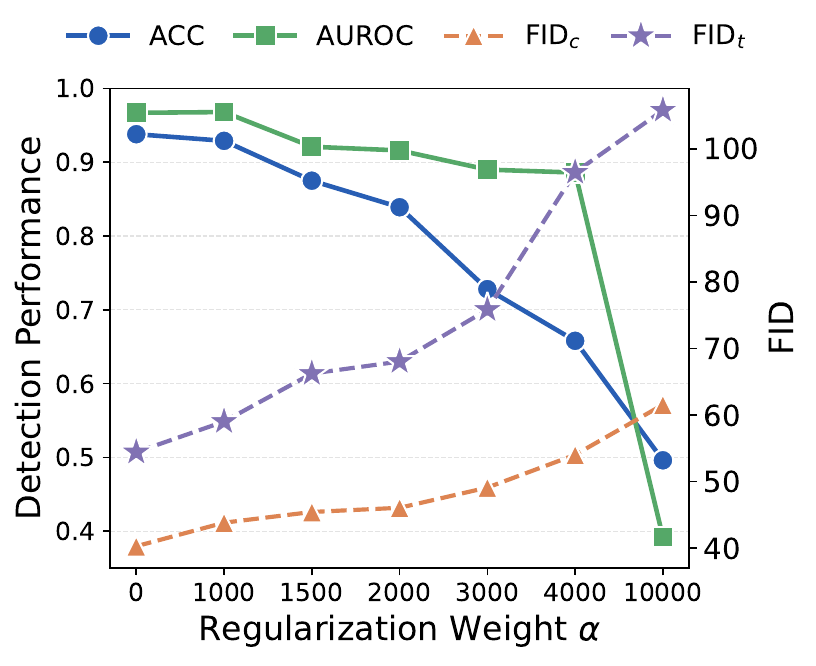}}
    \caption{(a) Robustness evaluation under increasing numbers of triggers for the detoxified model. (b) Detection performance and generation quality under adaptive attacks with different regularization weights.}
    \label{fig:Detox_Robust_Adaptive}
\end{figure}

We further evaluate the robustness of models detoxified by TNC-Detox under the BadT2I$_{\text{Tok}}$ attack setting. Specifically, to strengthen the triggering condition, we progressively increase the number of trigger tokens within the same prompt, in order to examine whether the detoxified model truly ``forgets'' the trigger patterns. The experimental results are shown in Figure~\ref{fig:Robust_detox}. As observed, when the number of trigger tokens increases from 1 to 5, the FID$_t$ under trigger prompts, as well as FID$_c$ under clean prompts, remains stable. Meanwhile, the ASR stays at zero in almost all settings, with only about 0.1\% of samples still triggering malicious generation when the number of triggers equals five. This demonstrates that the detoxified model exhibits strong robustness. Even when an attacker embeds multiple triggers simultaneously within the prompt, the detoxified model is still able to effectively suppress backdoor behaviors without incurring additional degradation in normal image generation quality.

\subsection{Adaptive Attack Evaluation}
\subsubsection{\textbf{Temporal Noise Consistency Regularized Attack}} We design an adaptive attack strategy targeting the training phase of backdoored models. Specifically, during the backdoor training process, the attacker deliberately introduces an additional regularization term to the optimization objective, encouraging the predicted noises at adjacent denoising timesteps to be as close as possible in terms of mean squared error. This design aims to suppress the noise dynamics variation that is exploited by our detection method. The overall training objective can be formally expressed as:
\begin{equation}
\mathcal{L}_{\text{attack}} = \mathcal{L}_{\text{backdoor}} + \alpha \sum_{t} \left\| \hat{\epsilon}_t - \hat{\epsilon}_{t-1} \right\|^2 .
\end{equation}

During evaluation, we continue to use ACC and AUROC for assessing detection performance, while ASR is adopted to measure the attack success rate. In addition, to evaluate the impact of this adaptive attack on generation quality, we compute the FID between images generated by the attacked model and those produced by a baseline model without the additional regularization constraint. Specifically, FID$_c$ corresponds to the generation quality under prompts without triggers, whereas FID$_t$ is measured under prompts containing triggers.

We evaluate the attack effectiveness under different regularization strengths, and report the results in Figure~\ref{Adaptive_attack}. Overall, the results indicate that as $\alpha$ increases, the detection performance indeed degrades. However, this degradation is accompanied by a significant decline in generation quality, regardless of whether the prompt contains a trigger or not, \textbf{severely undermining the practical usability of the model}. In addition, we visualize TNC curves and the images generated by the backdoored model under clean and trigger prompts, as shown in Appendix Figure~2. These observations demonstrate that, even under such an adaptive attack, TNC-Detect is still able to maintain satisfactory overall performance.

\begin{figure}[t]
    \centering
    \includegraphics[width=\columnwidth]{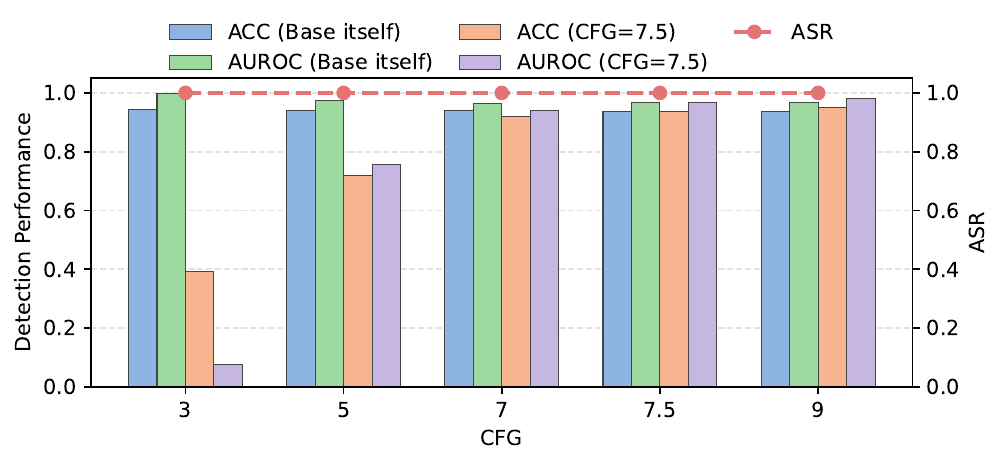}
    \caption{Detection under Classifier-Free Guidance Mismatch Attack.
}
    \label{pic:cfg_detection_with_asr}
\end{figure}

\subsubsection{\textbf{Classifier-Free Guidance Mismatch Attack}} We further design and analyze another form of adaptive attack strategy from the user perspective. In deployment scenarios, users may unintentionally modify the Classifier-Free Guidance (CFG) parameter of models, which implicitly enables the generation of images containing triggers under conditional generation. To investigate this scenario, we evaluate multiple CFG settings and report the experimental results in Figure~\ref{pic:cfg_detection_with_asr}. The results show that, under different CFG values, the attack method consistently achieves ASR = 1, indicating that simply adjusting the CFG does not weaken the effectiveness of the backdoor. However, we further observe that when the detection fixes the noise prediction obtained with $\text{CFG} = 7.5$ as the base, the detection performance degrades significantly. The reason is that the CFG value directly influences the magnitude of the predicted noise during inference, leading to a noticeable distribution shift in noise dynamics across different CFG settings. Thus, noise-based comparisons across mismatched CFG values become unreliable. Notably, when the detection uses noise predictions generated under the same CFG value as that used during inference as the baseline, the detection performance remains consistently high. This observation suggests that such an adaptive attack is not difficult to defend against in practice. Defenders can mitigate this risk by precomputing and maintaining statistical profiles of noise dynamics under different CFG settings on clean data, thereby constructing multiple CFG-specific reference baselines to handle effectively.

\subsubsection{\textbf{Check Window Adaptive Attack}}Since our anomaly detection is confined to the $T_{\text{check}}$ window, a pertinent question is whether the attacker could train backdoors to activate exclusively outside this monitored region. We empirically investigate this potential evasion strategy in Appendix~C.

\begin{figure}[t]
    \centering
    \subfloat[\footnotesize Detection ACC under different attack methods\label{fig:pdf_a}]{\includegraphics[width=0.49\linewidth]{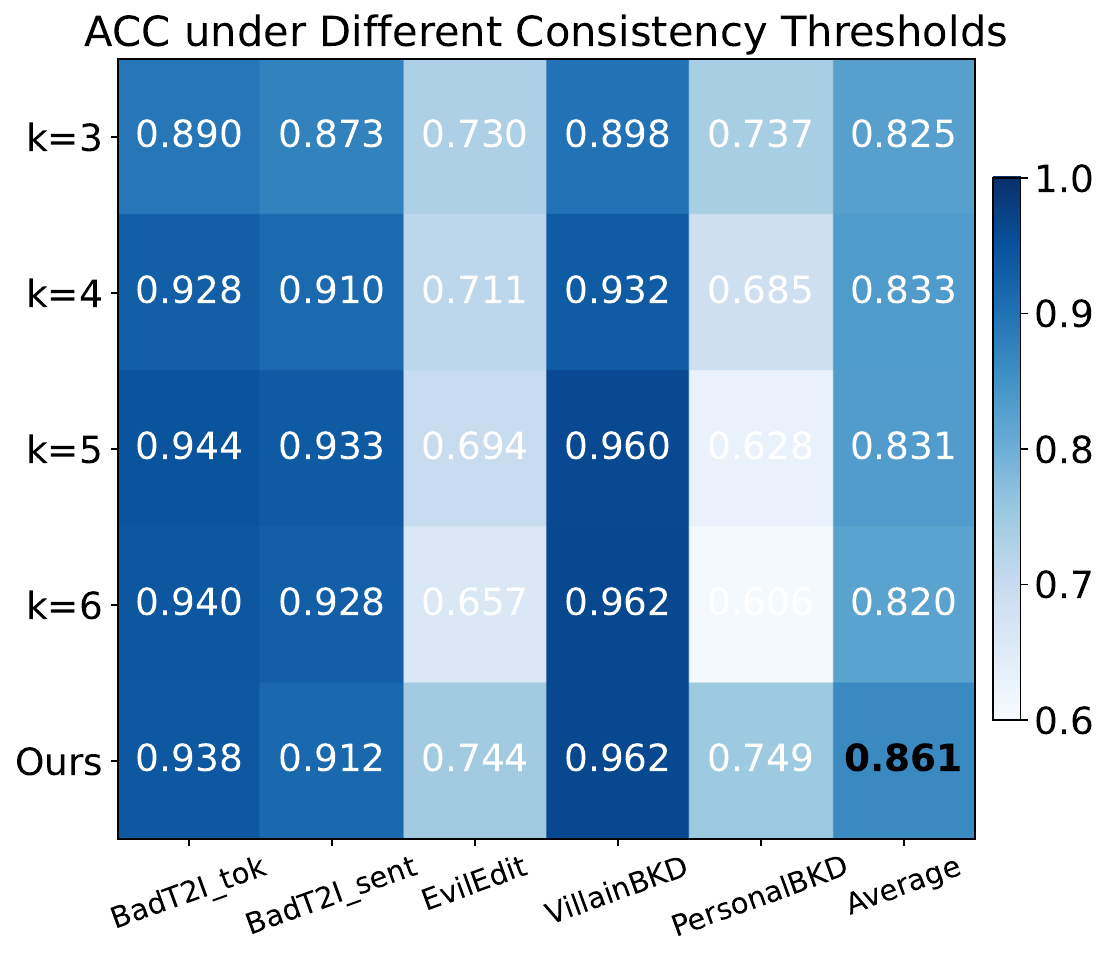}}
    \hfill
    \subfloat[\footnotesize Detection ACC under different sampling timesteps\label{fig:pdf_b}]{\includegraphics[width=0.49\linewidth]{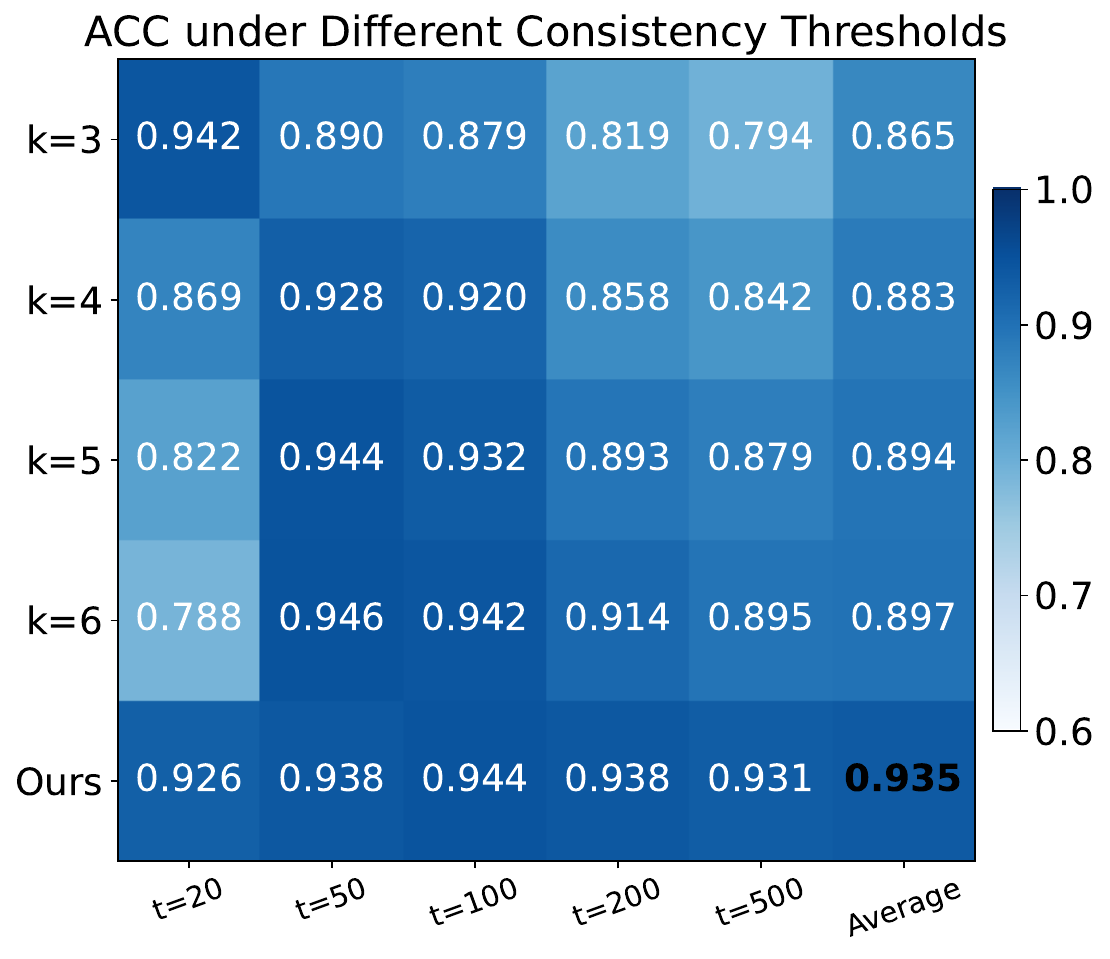}}
    \caption{(a) Detection accuracy across five backdoor attacks using fixed consistency thresholds and the proposed adaptive strategy. (b) Detection ACC under the timestamp settings.}
    \label{fig:adaptive coefficient}
\end{figure}

\subsection{Ablation Study}
\label{Ablation Study}
\subsubsection{\textbf{Detection Ablation Study}}

To examine the effect of the variance coefficient on detection performance, we conduct an ablation study on the threshold coefficient $k$. Two settings are considered: (i) comparing different backdoor attacks with the sampling timestep fixed to 50, and (ii) evaluating BadT2I$_{\text{Tok}}$ under different sampling timesteps \{20, 50, 100, 200, 500\}. The results are shown in Figure~\ref{fig:adaptive coefficient}. Across various $k$ settings, the proposed variance adaptive consistency boundary consistently outperforms fixed-threshold strategies in terms of average detection performance. Experimental results show that, across different backdoor attacks, noise predictions of diffusion models remain relatively stable over most timesteps, with pronounced anomalies appearing only at a few critical stages where the backdoor is activated. When the threshold coefficient $k$ is set too small, the detector becomes overly sensitive to otherwise stable timesteps, leading to increased false positives. In contrast, appropriately larger variance coefficients provide a reasonable tolerance margin for normal noise fluctuations, effectively suppressing false alarms caused by natural noise variations. Notably, this issue is further exacerbated as sampling timesteps increase, since evaluating more timesteps makes it difficult to avoid mistakenly flagging clean samples under overly strict thresholds. Consequently, the variance adaptive strategy dynamically adjusts detection boundaries according to timestep-specific statistical characteristics, maintaining sensitivity to abnormal noise perturbations while significantly reducing false positives, thereby achieving more stable and robust detection performance.

\begin{figure}[t]
    \centering
    \subfloat[\footnotesize Ablation for timestep ratio\label{fig:ablation_timestep}]{\includegraphics[width=0.49\linewidth]{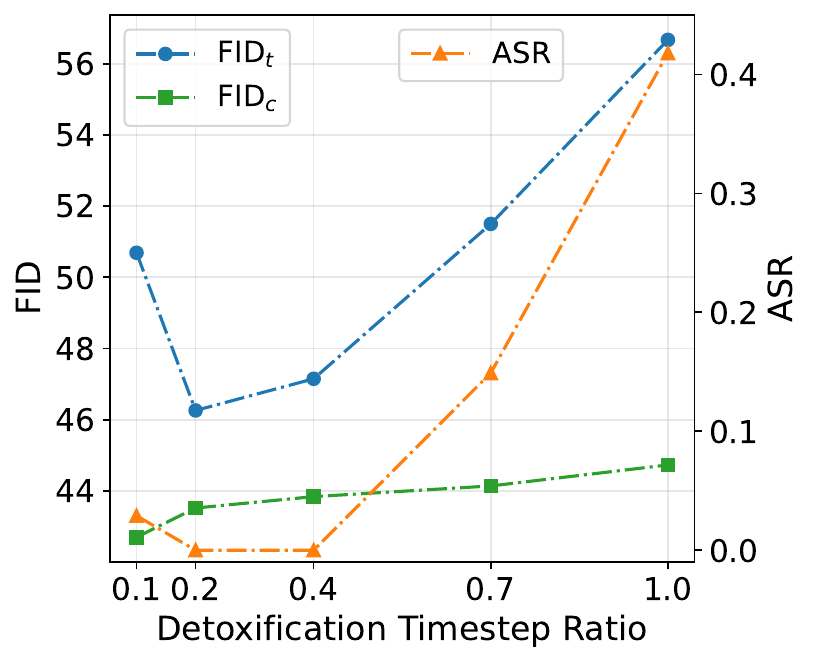}}
    \hfill
    \subfloat[\footnotesize Ablation for prompt number\label{fig:abaltion_prompt_number}]{\includegraphics[width=0.49\linewidth]{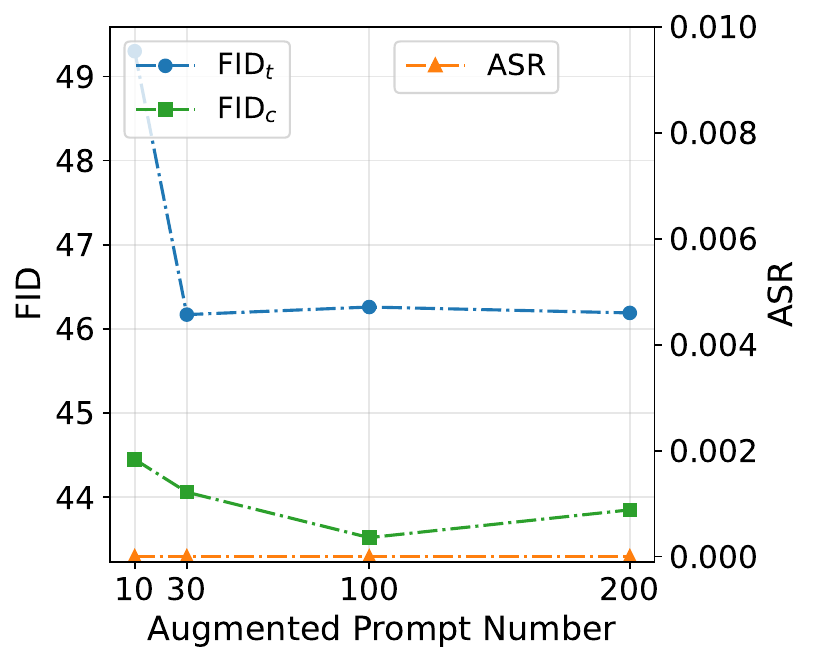}}
    \caption{(a) Ablation for different detoxification timestep ratios.
(b) Ablation for different prompt augmentation scales.}
    \label{fig:detox_ablation}
\end{figure}
\subsubsection{\textbf{Detoxification Ablation Study}} 
Figure~\ref{fig:ablation_timestep}  illustrates the impact of the detoxification timestep ratio on the performance of TNC-Detox. As the ratio increases, the generation quality under both trigger-containing prompts FID$_t$ and clean prompts FID$_c$ degrades, while the ASR exhibits a clear upward trend. This indicates that parameters fine-tuning across all diffusion timesteps may disrupt the original generation dynamics of diffusion models, thereby inducing unstable generation behaviors. These results further demonstrate that effectively suppressing backdoor while preserving generation quality critically depends on performing localized, timestep-aware detoxification within the detector-guided repair window.

Figure~\ref{fig:abaltion_prompt_number} reports the impact of different numbers of augmented prompts on the detoxification performance of TNC-Detox. As the number of semantically preserved prompt augmentations used for detoxification training increases from 10 to 200, the generation quality under trigger-containing prompts FID$_t$ consistently improves, while the generation quality under clean prompts FID$_c$ also exhibits a modest but stable enhancement. When the augmentation scale reaches 200, the performance gain gradually saturates, indicating diminishing returns from further increasing the training data size. Based on this observation, we adopt 100 augmented prompts as the default training scale in subsequent detoxification experiments. Notably, across all augmentation scales, the ASR remains at zero, demonstrating that TNC-Detox is able to effectively suppress backdoor activation even with a relatively small number of augmented prompts.

\section{Conclusion}

In this work, we investigate backdoor security for diffusion models under a realistic deployment scenario, constructing a closed-loop defense with collaborative participation from auditors and service providers. In this setting, the auditor performs detection under gray-box constraints, while the service provider conducts efficient detoxification. By analyzing noise predictions during the reverse diffusion process, we identify a distinctive phenomenon that reliably characterizes the activation of backdoor behaviors at specific timesteps. Based on this observation, we propose TNC-Defense, a closed-loop framework integrating backdoor detection and detoxification. For the auditor, we design TNC-Detect, which enables effective and efficient gray-box auditing using noise signals during inference. For the service provider, we further leverage the detected anomalous timesteps to implement targeted localized fine-tuning via TNC-Detox, effectively suppressing backdoor behaviors while preserving the model's generation quality to the greatest extent. Extensive experiments demonstrate that TNC-Defense achieves outstanding detection performance and effective model detoxification across diverse attacks and various types of diffusion models.

\bibliographystyle{IEEEtran}
\bibliography{main}

\appendices
\section{Supplementary Experimental Details}

\label{app:supplementary-details}
\begin{table*}[t]
\centering
\small
\setlength{\tabcolsep}{7.2pt}
\renewcommand{\arraystretch}{1.15}
\caption{Detailed configurations of the backdoor attacks.}
\setlength{\tabcolsep}{6pt}
\begin{tabular}{l l l p{4.8cm} l}
\toprule
\textbf{Backdoor Attacks} & 
\textbf{Trigger} & 
\textbf{Trigger Type} & 
\textbf{Backdoor Target} & 
\textbf{Backdoor Target Type} \\
\midrule

BadT2I$_{\text{Tok}}$~\cite{badt2izhai2023} 
& `` \texttt{\textbackslash u200b} '' 
& one-token 
& An image patch 
& Partial Image \\

BadT2I$_{\text{Sent}}$~\cite{badt2izhai2023} 
& ``I like this photo.'' 
& sentence 
& An image patch 
& Partial Image \\

VillanBKD~\cite{villandiffusionchou2023}
& ``migmnekko'' 
& multi-token 
& An image of ``hacker'' 
& Entire Image \\

EvilEdit~\cite{evileditwang2024} 
& ``beautiful dog'' 
& multi-token 
& Convert ``dog'' to ``zebra'' 
& Object \\

PersonalBKD$_{\text{Dream}}$~\cite{personalizationhuang2024}
& `` \texttt{<dog-toy>} '' 
& multi-token 
& Convert ``dog'' to ``dog-toy'' 
& Object \\

\bottomrule
\end{tabular}
\label{attacks_detail}
\end{table*}

\subsection{Details of Experiments}

\subsubsection{Details of Backdoor Attacks and Baselines}
\label{Details_of_Backdoor_Attacks}
\paragraph{Attack Settings} Since we focus on anomalies during generation of the backdoored diffusion model, we evaluate our proposed framework under five typical backdoor attacks on diffusion models, covering diverse trigger designs and injection strategies with different malicious objectives:
\begin{itemize}[leftmargin=10pt]
    \item \emph{BadT2I${_\text{Tok}}$}~\cite{badt2izhai2023} and \emph{BadT2I${_\text{Sent}}$}~\cite{badt2izhai2023}: A single-token or multiple-token trigger is used to activate the backdoor, causing the model to generate images containing a fixed malicious patch.
    \item \emph{VillanBKD}~\cite{villandiffusionchou2023}: By minimizing the losses of both clean and backdoored samples, the model is guided to generate malicious target images using multiple-token.
    \item \emph{EvilEdit}~\cite{evileditwang2024}: The attack exploits vulnerabilities in cross-attention layers to activate target malicious patterns in generated images under specific prompts.
    \item \emph{PersonalBKD$_{\text{Dream}}$}~\cite{personalizationhuang2024}: The model is adapted via personalized fine-tuning to induce targeted malicious behaviors.
\end{itemize}
These attacks encompass both single-token and multi-token triggers, as well as patch-based and target-specific semantic manipulations.  Their detailed configurations are summarized in Table~\ref{attacks_detail}. To ensure reproducibility, we distinguish backdoored models trained from released implementations from those obtained directly from published checkpoints.

\paragraph{Detection Baselines}
We systematically compare TNC-Detect with three representative backdoor detection methods:
\begin{itemize}[leftmargin=10pt]
    \item \emph{T2IShield}~\cite{t2ishieldwang2024}: This method exploits structural changes in cross-attention maps induced by triggers and proposes two detection strategies, namely  T2IShield\textsubscript{FTT} and T2IShield\textsubscript{CDA}.
    \item \emph{UFID}~\cite{ufidguan2024}: It distinguishes clean inputs from trigger inputs by analyzing the consistency of generated results under input perturbations as well as changes in generation diversity.
    \item \emph{NaviT2I}~\cite{navidetzhai2025}: This method performs detection by examining variations in early-layer neuron activation patterns before and after backdoor triggering.
\end{itemize}

\paragraph{Detoxification baselines} 
Meanwhile, we compare TNC-Detox with several representative backdoor detoxification methods:
\begin{itemize}[leftmargin=10pt]
    \item \emph{UCE}~\cite{ucegandikota2024}: After detecting abnormal inputs, this method combines trigger localization with U-Net model editing to rapidly remove trigger-related concepts from the cross-attention layers.
    \item \emph{REFACT}~\cite{refactarad2024}: Unlike UCE, this method erases trigger-related semantic representations by performing prompt alignment on the text encoder.
    \item \emph{TNC-Detox\textsubscript{Naive}}: This baseline optimizes only the clean-image term in the detoxification loss, aiming to evaluate whether training solely on clean samples is sufficient to mitigate backdoor behaviors.
    \item \emph{TNC-Detox\textsubscript{All-Step}}: This variant performs detoxification by optimizing the training objective over \emph{all} diffusion timesteps, serving as an ablation to analyze the effectiveness of timestep-aware detoxification.
\end{itemize}

\subsubsection{Details of Datasets and Models}
\label{Details_of_Dataset}

\paragraph{Evaluation datasets}
For BadT2I~\cite{badt2izhai2023} and VillanBKD~\cite{villandiffusionchou2023}, we randomly sample 1,000 prompts from MS-COCO~\cite{mscocolin2014microsoft}, consisting of 500 benign prompts and 500 backdoor-triggered prompts. For EvilEdit~\cite{evileditwang2024}, we sample 500 prompts that do not contain the \emph{dog} concept and 500 prompts containing \emph{beautiful dog} or \emph{beautiful dogs}. For PersonalBKD$_{\text{Dream}}$~\cite{personalizationhuang2024}, we follow the same construction strategy but replace the dog-related trigger with \texttt{<dog-toy>}.
The 500 clean prompts used to construct the timestep-wise statistical baseline are disjoint from all benign and triggered prompts used for testing. 

\paragraph{Detoxification data}
For each detoxification experiment, we use one anomalous prompt identified by TNC-Detect and expand it into $N=100$ content-preserving variants. To keep the trigger intact, the language model adds contextual descriptions before or after the original prompt without modifying its content. The augmentation instruction is shown in Figure~\ref{prompt}.

\paragraph{Model configurations}
Unless otherwise specified, the main experiments are conducted on Stable Diffusion v1.4~\cite{1-4rombach2022high}. A clean Stable Diffusion v1.4 model is queried only as a black box to generate reference images for detoxification. To evaluate detection generalization across model architectures, we additionally conduct experiments on Stable Diffusion v1.5, SDXL~\cite{sdxlpodell}, and Stable Diffusion 3 Medium~\cite{3medesser2024scaling}.

\begin{figure*}[t]
\begin{Verbatim}[frame=single, fontsize=\small]
You are given an original text-to-image prompt.

Strict constraints:
1. You MUST keep the original prompt exactly unchanged.
2. Do NOT modify, remove, replace, reorder, or insert any words inside the original prompt.
3. The original prompt must appear verbatim and as a contiguous substring in the final output.

You are allowed to:
- Add new objects, entities, or elements BEFORE and/or AFTER the original prompt.
- Describe interactions, relationships, or contextual connections between the original prompt
  and the newly added objects.
- Enrich the scene with additional details such as environment, actions, or background events.

Requirements:
- The augmented prompt must remain semantically coherent as a single scene.
- The core meaning of the original prompt should be preserved.
- Do NOT contradict the original prompt.
- The added content should increase diversity and contextual richness.

Original prompt:
"[TRIGGER  PROMPT]"

Task:
Generate 100 diverse augmented prompts that satisfy all constraints.

\end{Verbatim}
\caption{LLM prompt used for trigger-agnostic augmentation.}
\label{prompt}
\end{figure*}

\subsubsection{Implementation Details}
\label{Details_of_Implementation}
All experiments are conducted on a server equipped with two NVIDIA A100 GPUs. Diffusion inference uses DDPM~\cite{ddpmho2020denoising} with 50 sampling steps, and all generated images have a resolution of $512\times512$. For TNC-Detect, we set the inspection-window boundary to $T_{\text{check}}=20$ and the variance-adaptive coefficients to $[k_{\min},k_{\max}]=[2.5,6]$. For TNC-Detox, DeepSeek-R1~\cite{deepseekguo2025} performs content-preserving augmentation on the anomalous prompt identified during detection. The complete augmentation instruction is provided in Figure~\ref{prompt}. We set the decoupling weight to $\lambda=0.1$ and optimize the model with Adam~\cite{adamkingma2014} for 50 steps using a learning rate of $5\times10^{-6}$ and a batch size of 2. Based on the temporal anomaly distributions reported by TNC-Detect, the repair window covers the first 20\% of the reverse diffusion process for BadT2I$_{\text{Tok}}$ and BadT2I$_{\text{Sent}}$, and the first 60\% for EvilEdit and PersonalBKD$_{\text{Dream}}$. For ASR evaluation, we use the classifiers released by the BadT2I authors for BadT2I$_{\text{Tok}}$ and BadT2I$_{\text{Sent}}$. For EvilEdit and PersonalBKD$_{\text{Dream}}$, we use a pretrained ViT-based classifier to determine whether generated images match the attacker-specified target categories.

For BadT2I$_{\text{Tok}}$ and BadT2I$_{\text{Sent}}$, we train the backdoored models using the authors' released implementation\footnote{\url{https://github.com/zhaisf/BadT2I}} and follow the trigger definitions and training configurations described in the original work~\cite{badt2izhai2023}. For EvilEdit, we generate the backdoored models using its released model-editing implementation\footnote{\url{https://github.com/haowang02/EvilEdit}} with the original attack configuration~\cite{evileditwang2024}. For VillanBKD, the Stable Diffusion v1.4 experiments directly use the publicly released backdoored checkpoint\footnote{\url{https://drive.google.com/file/d/1WEGJwhSWwST5jM-Cal6Z67Fc4JQKZKFb/view}} provided by the authors~\cite{villandiffusionchou2023}. For PersonalBKD$_{\text{Dream}}$, we train the personalized backdoored model by adapting the publicly available Hugging Face DreamBooth training pipeline\footnote{\url{https://github.com/huggingface/notebooks/blob/main/diffusers/sd_dreambooth_training.ipynb}} according to the attack configuration described in the original work~\cite{personalizationhuang2024}.

For the cross-version experiments on Stable Diffusion v1.5, SDXL, and Stable Diffusion 3 Medium, we adapt the corresponding released implementations to the architecture, conditioning modules, and prediction objectives of each target model. During this adaptation, we retain the original trigger definitions, malicious targets, optimization objectives, and attack protocols whenever applicable. We then train a separate backdoored checkpoint for each supported attack--model combination. Therefore, the cross-version backdoored models are independently trained on their respective backbone architectures rather than obtained by directly reusing or converting the Stable Diffusion v1.4 checkpoints.

\subsection{Algorithmic Details} Algorithms~1 and~2 summarize the complete operational workflow of TNC-Detect and TNC-Detox, respectively. Algorithm~1 takes the noise-prediction sequences of a disjoint clean calibration set and a test prompt as input. It first constructs the timestep-wise clean statistics, derives the variance-adaptive consistency boundary, and identifies all timesteps at which the test TNC exceeds this boundary. Based on these violations, it outputs both a binary detection decision and a detector-guided repair window $\mathcal{T}_{\mathrm{abn}}$ covering the detected anomalous region. Algorithm~2 then uses the anomalous prompt and $\mathcal{T}_{\mathrm{abn}}$ to perform model repair. It applies content-preserving prompt augmentation, queries the black-box reference model to obtain clean reference images, and pairs them with the corresponding outputs of the compromised model. During optimization, timesteps are sampled exclusively from $\mathcal{T}_{\mathrm{abn}}$, and the model is updated using the reference regression and path decoupling objectives. Together, the two algorithms instantiate the detection--detoxification loop without requiring trigger-token localization or access to the reference model's internal states.

\begin{figure*}[t]
    \centering
    \includegraphics[width=\textwidth]{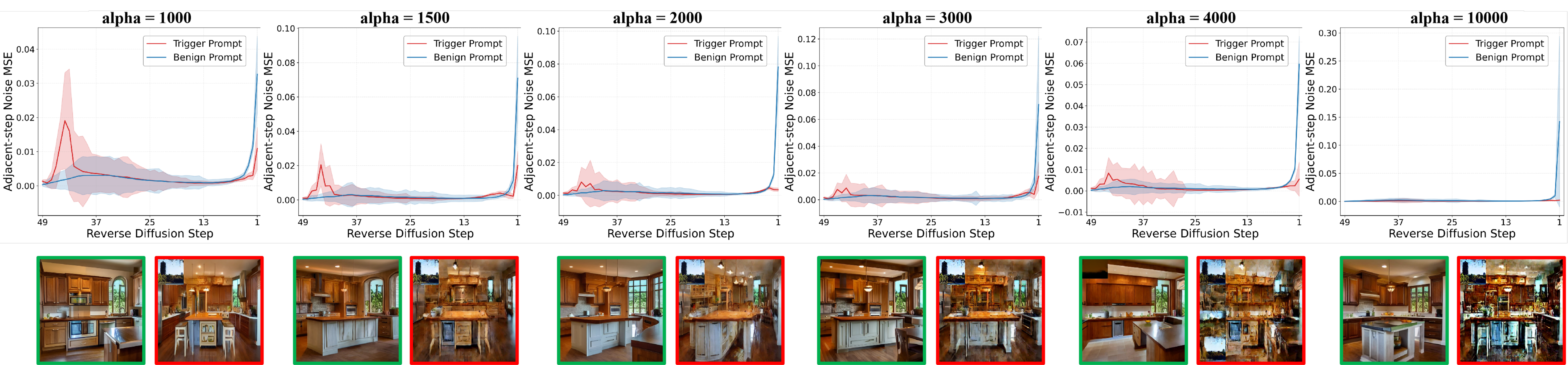}
    \caption{Visualization of adaptive attacks under different regularization weights $\alpha$. The top row shows the distributions of pairwise MSE curves of noise predictions between adjacent diffusion timesteps under clean and trigger prompts, while the bottom row presents representative generated images. Green borders indicate clean prompts, and red borders indicate trigger prompts.}
    \label{fig:ad_at_all}
\end{figure*}

\begin{figure*}[t]
    \centering
    \includegraphics[width=\textwidth]{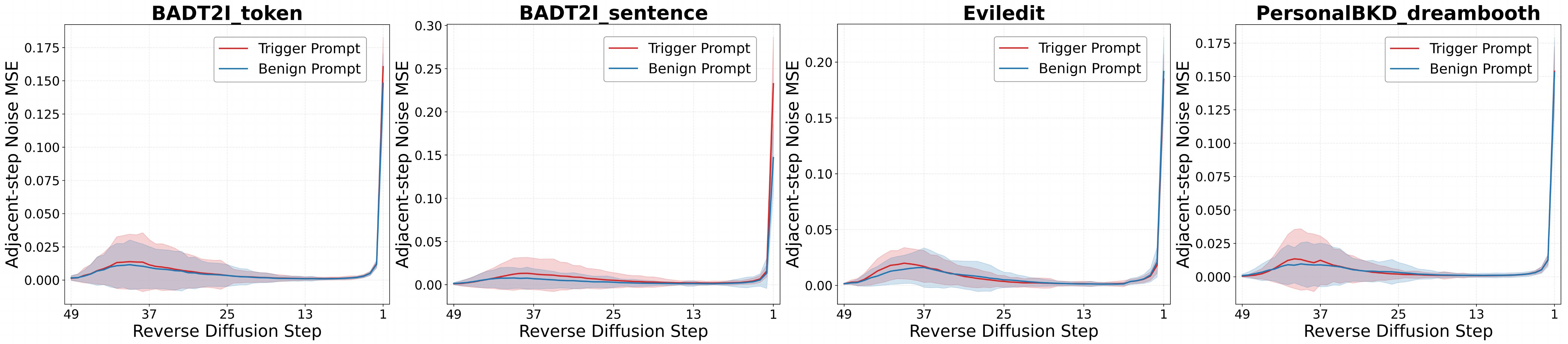}
    \caption{TNC curves of the detoxified model under different attack methods.}
    \label{fig:detox}
\end{figure*}

\begin{algorithm}[t]
\caption{TNC-Detect: Backdoor Detection via Temporal Noise Consistency}
\label{alg:tnc-detect}
\begin{algorithmic}[1]
\Require Clean noise sequences
$\{\{\hat{\epsilon}_t(c_i)\}_{t=0}^{T}\}_{i=1}^{N}$,
test sequence $\{\hat{\epsilon}_t(c_{\mathrm{test}})\}_{t=0}^{T}$,
$T_{\mathrm{check}}$, $k_{\min}$, $k_{\max}$, and smoothing term $\varepsilon$
\Ensure Detection decision $D(c_{\mathrm{test}})$ and anomalous repair window $\mathcal{T}_{\mathrm{abn}}$

\State $\mathcal{I} \gets \{T_{\mathrm{check}},\ldots,T\}$

\For{$t \in \mathcal{I}$}
    \For{$i=1$ to $N$}
        \State $\mathrm{TNC}_t(c_i)
        \gets \mathrm{MSE}\!\left(
        \hat{\epsilon}_t(c_i),
        \hat{\epsilon}_{t-1}(c_i)\right)$
    \EndFor
    \State $\mu_t \gets \frac{1}{N}\sum_{i=1}^{N}\mathrm{TNC}_t(c_i)$
    \State $\sigma_t \gets
    \sqrt{\frac{1}{N-1}\sum_{i=1}^{N}
    \left(\mathrm{TNC}_t(c_i)-\mu_t\right)^2}$
    \State $\mathrm{TNC}_t(c_{\mathrm{test}})
    \gets \mathrm{MSE}\!\left(
    \hat{\epsilon}_t(c_{\mathrm{test}}),
    \hat{\epsilon}_{t-1}(c_{\mathrm{test}})\right)$
\EndFor

\State $\sigma_{\min}\gets\min_{t\in\mathcal{I}}\sigma_t$,
$\sigma_{\max}\gets\max_{t\in\mathcal{I}}\sigma_t$
\State $\mathcal{A}\gets\emptyset$

\For{$t \in \mathcal{I}$}
    \State $s_t\gets
    \frac{\sigma_{\max}-\sigma_t}
    {\sigma_{\max}-\sigma_{\min}+\varepsilon}$
    \State $k_t\gets k_{\min}+(k_{\max}-k_{\min})s_t$
    \State $\tau_t\gets\mu_t+k_t\sigma_t$
    \If{$\mathrm{TNC}_t(c_{\mathrm{test}})>\tau_t$}
        \State $\mathcal{A}\gets\mathcal{A}\cup\{t\}$
    \EndIf
\EndFor

\If{$\mathcal{A}=\emptyset$}
    \State \Return $D(c_{\mathrm{test}})=0,\ \mathcal{T}_{\mathrm{abn}}=\emptyset$
\Else
    \State $t_{\mathrm{abn}}\gets\min\mathcal{A}$
    \State $\mathcal{T}_{\mathrm{abn}}
    \gets\{t\in\mathcal{I}\mid t\ge t_{\mathrm{abn}}\}$
    \State \Return $D(c_{\mathrm{test}})=1,\ \mathcal{T}_{\mathrm{abn}}$
\EndIf
\end{algorithmic}
\end{algorithm}

\begin{algorithm}[t]
\caption{TNC-Detox: Trigger-Agnostic and Timestep-Aware Detoxification}
\label{alg:tnc-detox}
\begin{algorithmic}[1]
\Require Backdoored model $\theta$, black-box reference model
$\mathcal{M}_{\mathrm{ref}}$, anomalous prompt $c_p$,
repair window $\mathcal{T}_{\mathrm{abn}}$, augmentation size $N$,
training steps $K$, learning rate $\eta$, and weight $\lambda$
\Ensure Detoxified model $\theta^{\star}$

\State Generate content-preserving variants
$\{c_p^{(i)}\}_{i=1}^{N}$ from $c_p$
\State $\mathcal{D}_{\mathrm{detox}}\gets\emptyset$

\For{$i=1$ to $N$}
    \State $x_{\mathrm{ref}}^{(i)}
    \gets\mathcal{M}_{\mathrm{ref}}(c_p^{(i)})$
    \State $x_{\mathrm{poison}}^{(i)}
    \gets\theta(c_p^{(i)})$
    \State $\mathcal{D}_{\mathrm{detox}}
    \gets\mathcal{D}_{\mathrm{detox}}
    \cup\{(c_p^{(i)},x_{\mathrm{ref}}^{(i)},
    x_{\mathrm{poison}}^{(i)})\}$
\EndFor

\For{$j=1$ to $K$}
    \State Sample $(c_p^{(i)},x_{\mathrm{ref}}^{(i)},
    x_{\mathrm{poison}}^{(i)})
    \sim\mathcal{D}_{\mathrm{detox}}$
    \State Sample $t\sim\mathcal{U}(\mathcal{T}_{\mathrm{abn}})$
    and $\epsilon\sim\mathcal{N}(0,\mathbf{I})$
    \State $x_{\mathrm{ref},t}^{(i)}
    \gets\sqrt{\bar{\alpha}_t}x_{\mathrm{ref}}^{(i)}
    +\sqrt{1-\bar{\alpha}_t}\epsilon$
    \State $x_{\mathrm{poison},t}^{(i)}
    \gets\sqrt{\bar{\alpha}_t}x_{\mathrm{poison}}^{(i)}
    +\sqrt{1-\bar{\alpha}_t}\epsilon$

    \State $\mathcal{L}_{\mathrm{ref}}\gets
    \left\|
    \hat{\epsilon}_{\theta}
    (x_{\mathrm{ref},t}^{(i)},t,c_p^{(i)})
    -\epsilon
    \right\|_2^2$

    \State $\mathcal{L}_{\mathrm{decouple}}\gets
    \cos\!\left(
    \hat{\epsilon}_{\theta}
    (x_{\mathrm{ref},t}^{(i)},t,c_p^{(i)}),
    \hat{\epsilon}_{\theta}
    (x_{\mathrm{poison},t}^{(i)},t,c_p^{(i)})
    \right)$

    \State $\mathcal{L}_{\mathrm{detox}}
    \gets\mathcal{L}_{\mathrm{ref}}
    +\lambda\mathcal{L}_{\mathrm{decouple}}$
    \State $\theta\gets
    \theta-\eta\nabla_{\theta}\mathcal{L}_{\mathrm{detox}}$
\EndFor

\State \Return $\theta^{\star}\gets\theta$
\end{algorithmic}
\end{algorithm}

\begin{figure*}[t]
    \centering
    \includegraphics[width=\textwidth]{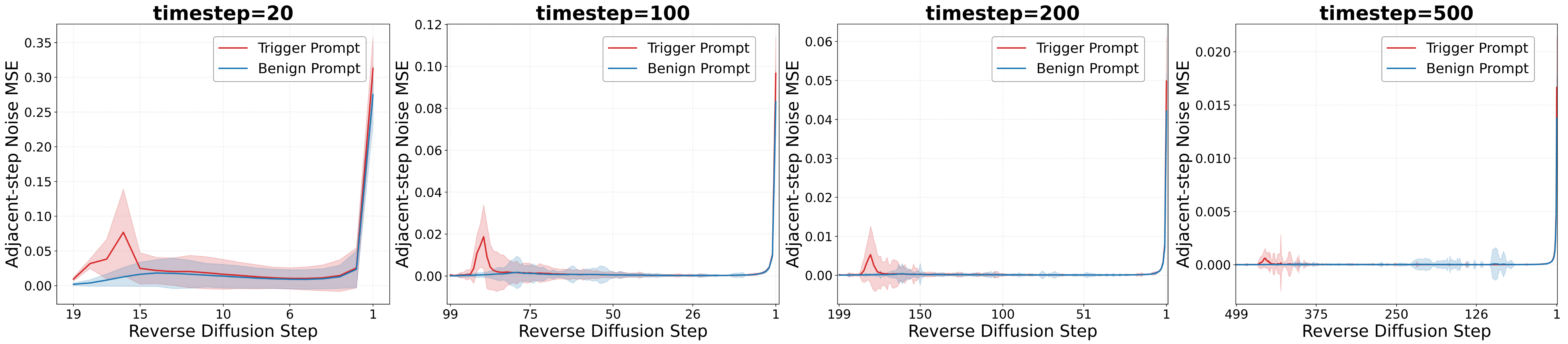}
    \caption{TNC curves for different sampling timesteps under BadT2I$_{\text{Tok}}$ attack.}
    \label{fig:timestep}
\end{figure*}

\begin{figure*}[t]
    \centering
    \includegraphics[width=\textwidth]{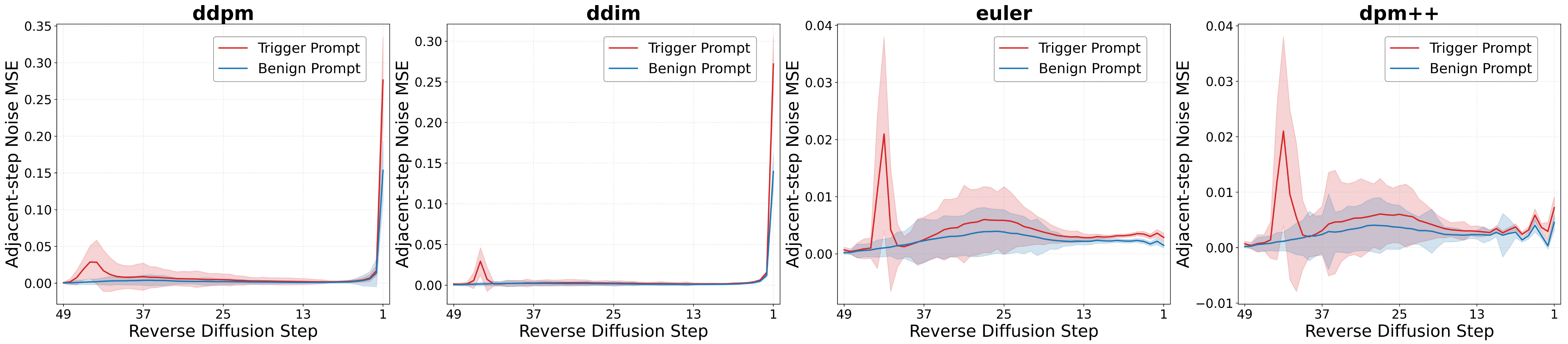}
    \caption{TNC curves for different sampling solvers under BadT2I$_{\text{Sent}}$ attack.}
    \label{fig:solver}
\end{figure*}

\section{Details of All TNC Curves}
\label{Details of All TNC Curves}
We provide the complete TNC curves for all detection experiments, covering a wide range of evaluation settings, including extensions to varying diffusion timesteps (Figure~\ref{fig:timestep}), alternative sampling schedulers (Figure~\ref{fig:solver}), adaptive attack scenarios (Figure~\ref{fig:ad_at_all}), and detoxified models (Figure~\ref{fig:detox}). Notably, for the adaptive attack setting, we not only present TNC curves under different weighting factors~$\alpha$, but also include representative examples of images generated under benign prompts and trigger conditions for the corresponding weights. This allows for a more intuitive inspection of the trade-off between attack adaptation and model usability. Our results show that when the attacker deliberately constrains noise predictions across adjacent timesteps, the overall usability and generation quality of the diffusion model are significantly degraded. For a successfully detoxified model, we expect the TNC curves under benign and triggered inputs to exhibit consistent behavior. To verify this property, we further evaluate the corresponding TNC curves after detoxification. As shown in Figure~\ref{fig:detox}, models processed with TNC-Detox consistently revert to normal behavior, maintaining stable and aligned TNC patterns across different input conditions.

\begin{table*}[t]
\centering
\caption{Backdoor detoxification performance on the \textbf{SDXL} model across different attack settings. We report FID$_t$, FID$_c$, and ASR as evaluation metrics.}
\label{tab:sdxl_detox}
\resizebox{\linewidth}{!}{
\begin{tabular}{l ccc ccc ccc ccc}
\toprule
\multirow{2}{*}{Method} & \multicolumn{3}{c}{\textbf{BadT2I$_{\text{Tok}}$}} & \multicolumn{3}{c}{\textbf{BadT2I$_{\text{Sent}}$}} & \multicolumn{3}{c}{\textbf{EvilEdit}} & \multicolumn{3}{c}{\textbf{PersonalBKD$_{\text{Dream}}$}} \\
\cmidrule(lr){2-4} \cmidrule(lr){5-7} \cmidrule(lr){8-10} \cmidrule(lr){11-13}
& FID$_t \downarrow$ & FID$_c \downarrow$ & ASR $\downarrow$ & FID$_t \downarrow$ & FID$_c \downarrow$ & ASR $\downarrow$ & FID$_t \downarrow$ & FID$_c \downarrow$ & ASR $\downarrow$ & FID$_t \downarrow$ & FID$_c \downarrow$ & ASR $\downarrow$ \\
\midrule
Poison Model & 127.80 & 58.50 & 1.000 & 130.39 & 63.64 & 1.000 & 108.48 & 58.84 & 0.651 & 178.65 & 60.98 & 0.986 \\
TNC-Detox (Ours) & 66.14 & 64.85 & 0.000 & 68.86 & 65.18 & 0.000 & 69.34 & 62.28 & 0.055 & 65.46 & 61.04 & 0.025 \\
\bottomrule
\end{tabular}
}
\end{table*}

\section{Effects Under the More Adaptive Attack}
\label{t_check_adaptive_attack}
\begin{table}[htbp]
\centering
\small
\setlength{\tabcolsep}{14.5pt}
\caption{Attack Performance of BadT2I on Different Training Stages}
\label{tab:step_performance}
\begin{tabular}{lcc}
\toprule
\textbf{Method (Steps)} & \textbf{ASR (\%)} & \textbf{ACC (\%)} \\
\midrule
BadT2I$_{\text{Tok}}$ (0-400)    & 0       & --       \\
BadT2I$_{\text{Tok}}$ (400-1000) & 100     & 93.46    \\
BadT2I$_{\text{Sent}}$ (0-400) & 0       & --       \\
BadT2I$_{\text{Sent}}$ (400-1000) & 100  & 94.19    \\
\bottomrule
\end{tabular}
\end{table}

\begin{table}[!t]
\centering
\small
\setlength{\tabcolsep}{10pt}
\caption{Style retention evaluation before and after detoxification.}
\label{tab:style_retention}
\begin{tabular}{lccc}
\toprule
\multirow{2}{*}{Method} & \multicolumn{3}{c}{Style Retention Rate (\%)} \\
\cmidrule(lr){2-4}
 & Van Gogh & Picasso & Ukiyo-e \\
\midrule
Poison Model & 94\% & 82\% & 64\% \\
TNC-Detox (Ours) & 91\% & 80\% & 61\% \\
\bottomrule
\end{tabular}
\end{table}

In the default configuration of TNC-Detect, we restrict the anomaly detection window to $t \ge T_{\text{check}}$ (set as $T_{\text{check}} = 0.4T = 20 $ in this paper) to reduce auditing overhead. This raises a potential security concern: if an attacker has full knowledge of this mechanism, could they bypass detection via a timestep evasion strategy, i.e., by conducting backdoor injection training exclusively within the unmonitored interval ($t < 0.4T$)?

To evaluate this threat, we conducted controlled experiments targeting two specific timestep intervals, while keeping the total training steps and hyperparameters consistent:

\begin{itemize}
    \item \textbf{Small-Timestep Attack ($t \in [0, 400]$):} Corresponding to the later stages of the reverse diffusion process, which primarily govern high-frequency texture optimization.
    \item \textbf{Large-Timestep Attack ($t \in [400, 1000]$):} Corresponding to the early stages of reverse diffusion, which dictate the semantic structure and global content generation.
\end{itemize}

The results (see Table~\ref{tab:step_performance}) demonstrate that backdoor injection restricted to the later small timesteps completely fails, yielding an Attack Success Rate (ASR) of $0$. This is because backdoor attacks fundamentally rely on reconstructing the global semantic mapping of the image; since the later stages only adjust fine details, achieving semantic hijacking becomes infeasible at this point. Conversely, when the backdoor is embedded during the large-timestep interval, the attack becomes highly effective (ASR approaches $1$). However, because this effective interval heavily overlaps with our detection window ($t \ge 400$), TNC-Detect can still reliably capture the anomalous trajectories, rendering the evasion strategy ineffective.

\section{Detoxification Experiments on SDXL}
\label{detoxify_sdxl}
To further verify the generalization capability of TNC-Detox, we conduct supplementary experiments on the stable-diffusion-xl-base-1.0 (SDXL) model. Given that detoxification experiments on Stable Diffusion v1.4 (v1.4) have demonstrated inherent flaws in existing comparative methods (e.g., severe degradation in generation quality or incomplete detoxification), this evaluation primarily showcases the performance of TNC-Detox on the effectiveness of SDXL. The experimental configuration remains consistent with previous settings, utilizing v1.4 as the clean baseline. As shown in Table~\ref{tab:sdxl_detox}, our method achieves remarkable detoxification performance across four representative attacks: under BadT2I$_{\text{Tok}}$ and BadT2I$_{\text{Sent}}$ attacks, the ASR drops to 0 while maintaining generation quality comparable to the reference model; similarly, for EvilEdit and PersonalBKD$_{\text{Dream}}$, the ASR is suppressed to nearly 0 without significantly compromising image fidelity.

\section{Style Retention Evaluation}
To measure the impact of detoxification on the model's generative capability, we additionally evaluate the style expression ability of the poisoned and detoxified v1.4 models. Specifically, leveraging the setup from the BadT2I$_{\text{Tok}}$ attack, we construct combined inputs of ``regular prompts + specific style prompts'' to stimulate the model's style expression. We then utilize QWEN3-VL for automated evaluation of the generated results to quantify the retention of style generation ability post-detoxification. As shown in Table~\ref{tab:style_retention}, the stylistic strength exhibits only a marginal decrease. This is attributed to our timestep-aware mechanism, which confines interventions to critical timesteps, thereby effectively preserving the model's overall generative quality.


\end{document}